\documentclass{aa-jun06}   
\usepackage{graphicx}             
\usepackage{natbib}
\usepackage{subfigure}
\usepackage{ulem}
\usepackage{dcolumn}
\usepackage{txfonts}

\include{definitions}

      \def\new#1 {\bf {#1 }}
      \def\cut#1 {\sout {#1 }}

\def\kms {$\mathrm{km\,s^{-1}}$} 
\def\AMM {$\mathrm{NH_3}$} 
\def\DAMM {$\mathrm{NH_{2}D}$} 
\def\DDAMM {$\mathrm{NHD_2}$} 
\def\CO {$\mathrm{C^{18}O}$} 
\def\COs {$\mathrm{C^{17}O}$} 
\def\htcn {$\mathrm{H^{13}CN}$} 
\def\hcfn {$\mathrm{HC^{15}N}$} 
\def\percc {$\mathrm{cm^{-3}}$} 

\def\simgreat{\mathbin{\lower 3pt\hbox
     {$\rlap{\raise 5pt\hbox{$\char'076$}}\mathchar"7218$}}}
\def\simless{\mathbin{\lower 3pt\hbox
     {$\rlap{\raise 5pt\hbox{$\char'074$}}\mathchar"7218$}}} 
\def\tdv {$\int T_{\rm MB}\rm d$v} 

\begin{document}

\title{ Probing the initial conditions of High Mass Star formation -- ${\rm I}$.}
 \subtitle{Deuteration and depletion in high mass pre/protocluster clumps}
 %

  \author{T. Pillai \inst{1}\thanks{Present address: Harvard-Smithsonian Centre for Astrophysics},  F. Wyrowski
          \inst{1},
            J.Hatchell \inst{2},  A.G. Gibb \inst{3}
          \and
           M.A. Thompson \inst{4}
         }

   \offprints{tpillai@cfa.harvard.edu}

   \institute{Max-Planck-Institut f\"{u}r Radioastronomie, Auf dem H\"{u}gel 69, D-53121 Bonn, Germany \\
              \email{tpillai@cfa.harvard.edu}
               \and
               School of Physics, University of Exeter, Stocker Road, Exeter EX4
4QL, UK\\
               \and
               Department of Physics \& Astronomy, University of British Columbia, Vancouver, BC, V6T
1Z1, Canada \\
               \and
               Centre for Astrophysics Research, Science \& Technology Research Institute, University
of Hertfordshire, College Lane, Hatfield, AL10 9AB, UK \\
              }

 \abstract
{}
{UltraCompact H{\sc ii} regions are signposts of high mass star
formation.  Since high-mass star formation occurs in clusters, one
expects to find even earlier phases of massive star formation in the
vicinity of UltraCompact H{\sc ii} regions.  Here, we study
the amount of deuteration and depletion toward pre/protocluster clumps found in a wide-field ($10\times 10$~arcmin)
census of clouds in $32$ massive star-forming regions that are known
to harbour UCH{\sc ii} regions.}
{We determine the column density of \AMM, \DAMM, CO, \htcn, and \hcfn\
lines. We used the $({J,K}) = (1,1)$ and (2,2) inversion transitions
of \AMM\ to constrain the gas temperatures.}
 {We find that 65\% of the observed sources have strong \DAMM\
emission and more than 50\% of the sources exhibit a high degree of
deuteration, ($0.1\le \mathrm{NH_{2}D/NH_3} \le$ 0.7), 0.7 being the
highest observed deuteration of \AMM\ reported to date. Our search for \DDAMM\ in two sources did not result in a detection. The
enhancement in deuteration coincides with moderate CO depletion onto
dust grains. There is no evidence of a correlation between the two
processes, though an underlying correlation cannot be ruled out as the
depletion factor is very likely to be only a lower limit. Based on
simultaneously observed \htcn\ and \hcfn\ ($J$=1--0) lines, we derive
a high abundance ratio of \htcn\ to \hcfn, which might indicate
anomalous ratios of C and N isotopes relative to those derived toward
the local ISM.}
{We find CO depletion and high deuteration towards cold cores in
massive star forming regions. Therefore, these are good candidates for
sources at the early phases of massive star formation. While our
sensitive upper limits on \DDAMM\ do not prove the predictions of
the gas-phase and grain chemistry models wrong, an enhancement of $\approx 10^4$
over the cosmic D/H ratio from \DAMM\ warrants explanation.}

\authorrunning{Pillai et al.\ 2007}
\titlerunning{Deuteration and Depletion in High Mass Clumps}
\maketitle

\section{Introduction}

The earliest phases of high mass star formation are still poorly
understood.  The prevalent theory is that massive stars are born in
dense clumps within giant molecular clouds (GMCs), where the presence
of Ultracompact HII regions (UCHII) identifies massive stars which
have already begun to ionize their surroundings. Observations
\citep{Cesaroni1994,Cesaroni1998} have shown that these
are often associated with warm ($T>100$~K), compact($<0.1$~pc) and
very dense ($n_{\rm H_2}>10^7$~\percc) cores known as hot molecular
cores (HMCs). An evolutionary sequence has been suggested based on
recent observations (\citealt{codella2004},
\citealt{beuther2005:astroph_review}) with the HMC stage immediately
preceding the formation of UCH{\sc ii} regions.  But the stage before hot cores --
the precluster and early protocluster phases of massive star formation
-- have not yet been studied. Since many UCH{\sc ii} regions are
located in clusters, one expects to find even earlier phases of
massive star formation, and the raw material out of which massive stars
or star clusters form, in the vicinity of UCH{\sc ii}s
\citep{thompson2006a:scuba_uchii}.

This has prompted us to embark upon a program to search for massive
pre/protocluster clumps by taking a wide-field ($10\times
10$~arcmin) census of clouds in $32$ massive star-forming
regions, harbouring UCH{\sc ii} regions \citep{wood1989:uchii}. Our program is known as SCAMPS (the
\underline{SC}UB\underline{A} \underline{M}assive
\underline{P}re/Protocluster core \underline{S}urvey; \citealt{thompson2005:dmuconf}, Thompson et al.\ 2007 in prep.).

We discovered a multitude of secondary, cold condensations and warmer
clumps that might contain heavily embedded massive protostars. Many of
the sources are seen as dark patches in MSX infrared images of the
region, infrared dark clouds (IRDCs;
\citealt{menten2005:iau}). As a result these clumps must have dust
temperatures below 30~K (as evidenced by MIR upper limits) and
have masses of a few 100 to 1000 M$_\odot$.  These clumps thus could
be in a colder pre-protostellar phase.

In order to study the physical and chemical conditions of these
clumps, we carried out a multi-wavelength survey toward them in
various molecular tracers. In this study, we also included 9 sources
which were selected on the basis of the MIR extinction (IRDCs) and
followed up later with submillimeter dust continuum emission and
millimeter rotational lines of H$_2$CO \citep{carey1998:irdc}. We have
recently reported a study of ammonia in this sample
\citep{pillai2006b:nh3}. Here, we report on our study using ammonia to
probe the temperature and deuteration in the clumps and CO to estimate
depletion of neutral molecules from the gas phase. Depletion has
proved to be a good tracer of low-mass pre-stellar cores
(\citealt{bacmann2003:depln} and
\citealt{crapsi2005:deutn_depn}), and \AMM\ is an important
tool in measuring the physical conditions in molecular clouds and can
be used to confirm the low temperatures and high densities required
for depletion \citep{ho1983:nh3}. Since only the lowest \AMM\ energy
levels are expected to be populated for cold dark clouds ($T<20$~K),
their physical conditions can be probed using the (1,1) and (2,2)
inversion transitions in the metastable ($J,K$) rotational levels of
ammonia. Radiative transitions between different $K$-ladders are
forbidden, therefore the lowest levels are populated only via
collisions. The optical depth can be determined from the ratio of the
hyperfine satellites. Thus, the population of the different levels can
be estimated and hence the temperature of the gas determined.

\AMM\ is observed to be an excellent tracer
of the dense gas where many other molecules would have heavily
depleted \citep{tafalla2004:starless_cores}. In addition, recent chemical models reveal that \AMM\
(and also N$_2$H$^+$), does not deplete from the gas phase for the
densities observed in dark clouds ($<10^{6}$~\percc; \citealt{bergin1997:chemistry}). However in the light of recent experiments  of the desorption of CO and N$_2$ at temperatures relevant to dense cores, the relative difference between the derived CO and N$_2$ binding energies is found to be significantly less than that currently adopted in astrochemical models \citep{oeberg2005:co_n2}. Therefore, the observed 
non-depletion of \AMM\ and N$_2$H$^+$ during the early formation of a massive cluster remains a ``mystery''.

High densities in the earliest phase are expected to enhance
depletion of molecules (mainly CO) onto grains
\citep{bacmann2003:depln}. The freeze out of abundant gas phase
molecules along with the low temperatures ($\le$~20~K) prevalent in
these clumps leads to a high degree of deuteration
(\citealt{flower2005_freeze_out}, \citealt{roueff2005:deutn},
\citealt{hatchell2003:nh2d}, \citealt{shah2001:nh2d},
\citealt{saito2000:nh2d}). Hence, the estimation of deuteration and
depletion in clumps can be used as a chemical and physical
chronometer.

In this paper we investigate the \AMM\ temperature and deuteration in
our candidate pre/protocluster cores using observations with the
Effelsberg 100m and IRAM 30m telescopes. 
In Sect. \ref{sec:observations}, we describe our single dish
observations.
We also discuss the data reduction and present the derived physical
parameters for different molecules. In Sect.~\ref{sec:analysis}, we
show that candidate pre/protocluster clumps exhibit very high deuteration and
depletion. The fractionation of \AMM\ is compared with depletion of CO
from the gas phase. The possible origin of deuterated species and the
variation of deuteration with gas temperature is also discussed. 
HCN was also observed as part of the 30m observations and we
conclude the discussion with our
interpretation of the HCN isotopic abundance with galactic distance.


\section{Observations \& Results \label{sec:observations}}

In this section we present the results of single dish observations of
\AMM, \DAMM, CO and HCN lines on a sample of 23 SCAMPS sources and 9
IRDCs. The brightest compact submm sources with no MIR counterpart
were selected for these observations from our original catalogue of
the SCUBA fields. The parameters of the molecular transitions covered,
along with the observing setups are given in Table~\ref{tab:freq_list}. In
Table~\ref{tab:source_list}, we give the source list.  The derived
line parameters and the column densities are presented in Tables~3 --
8.

\subsection{IRAM 30m observations}
The \DAMM, \CO, \htcn\ and \hcfn\ observations were made at the IRAM 30m telescope \footnote{IRAM is an international institute for research in millimeter astronomy. IRAM is supported by INSU/CNRS
(France), MPG (Germany) and IGN (Spain).} on Pico Veleta in August 2003 toward the 32 candidate pre/protocluster
clumps given in Table~\ref{tab:source_list}. We performed 9 point maps
with 10$''$ spacing around the peak to be able to compare,
independently of the beam, lines at different frequencies. All line
parameters listed are averaged over the map.

Individual positions were
observed with an integration time of 2 minutes per position in the
position switch mode. The receivers A100, B100, A230 and B230 were
tuned to 86.13, 109.975, 219.56 and 231.32~GHz respectively.

The VESPA autocorrelator was used at the backend, with
a spectral resolution of 40~kHz and 320~kHz at 100 and 230~GHz
respectively. We had average summer weather conditions with a maximum
system temperature of 248~K at 110 GHz. Towards selected sources we also have \COs\
($J=2$--1) and \DDAMM\ (1$_{10}$--1$_{01}$)  observations.  The half power beam width (HPBW) of
the 30m telescope is 22$''$ at 110 GHz and 11$''$ at 230 GHz. The main
beam efficiency at these frequencies is 0.75 and 0.52, respectively.
\begin{table*}[h]
\begin{center}
\caption{Parameters of observed rotational transitions} \label{tab:freq_list}
\vspace{1em}

\begin{tabular}{lrccc}
\hline
\hline
 Species & Transition       & $E_{\rm l}^{\rm a}$ (K) & $\nu$ (MHz) & \\
\hline
\AMM\          & $(J,K=1,1)$              &    22.70        & 23694.496    \\
\AMM\          & $(J,K=2,2)$              &    63.89        & 23722.633    \\
 \DAMM\ (para) & 1$_{11}$--1$_{01}$       &    16.55        & 85926.3      \\
 \DAMM\ (ortho)& 1$_{11}$--1$_{01}$       &    15.98        & 110153.6     \\
\DDAMM\ (ortho)& 1$_{10}$--1$_{01}$       &    13.33        & 110812.9      \\
\DDAMM\ (para) & 1$_{10}$--1$_{01}$       &    13.09        & 110896.7     \\
 \CO\          & $J=1$--0                 &    0.0          & 109782.1734  \\
 \CO\          & $J=2$--1                 &    5.27         & 219560.3568  \\
 \COs\         & $J=2$--1                 &    5.39         & 224714.3850  \\
 \htcn\        & $J=1$--0, $F=2$--1       &    0.0          & 86340.184    \\
 \hcfn\        & $J=1$--0                 &    0.0          & 86054.961    \\
\hline
\end{tabular}
\end{center}
Note: $^{\rm a} \rm E_{\rm l}$ is the lower energy level of the transition.
\end{table*}

\subsection{Effelsberg 100m observations}

We observed the \AMM\ (1,1) and (2,2) transitions with the Effelsberg
100m telescope \footnote{Based on observations with the 100m telescope of the
MPIfR (Max-Planck-Institut für Radioastronomie) at Effelsberg).} in April 2004 for the 23 sources listed in Table~\ref{tab:source_list}. 
Two of the sources were re-observed because of baseline problems with the receiver.
With the AK 8192 backend, we were able to observe the (1,1), (2,2),
(3,3) and (4,4) transitions in both polarisations simultaneously using
the K-band receiver.  With 8 subunits of 10~MHz bandwidth, the
resulting spectral resolution was $\approx 0.25$~km$\,$s$^{-1}$.  The
beam at the \AMM\ frequencies was 40$''$.  The observations were
performed in the frequency switching mode.  Pointing was checked at
hourly intervals by continuum scans on G10.62.  We estimate the
pointing to be accurate to within 6~$''$.  The pointing scans were
used for the absolute flux calibration.  The \AMM\ observations toward the
9 IRDCs are reported in \citet{pillai2006b:nh3}. Note that for
G12.19-0.12 and G29.97-0.05 we report the observations carried out in
October 2002 with a spectral resolution of $\approx 1$~km$\,$s$^{-1}$.

\begin{table*}
\caption{Positions and velocities of the observed sources. IRDC positions are taken from Carey et al. 1998 }
\vspace{1em}
\begin{tabular}{rccc}
\hline
\hline
{Source} & { R.A.(2000)}& {Dec.(2000)}    & { $v_{\rm LSR}$ [km/s]}   \\
\hline


         &   SCAMPS     &                 &                       \\
\hline
 G8.13+0.25         & 18:02:55.69       & -21:47:46.7   & 19.4      \\
 G8.68-0.37         & 18:06:23.24       & -21:37:14.1   & 35.2/38.1 \\
 G8.71-0.37         & 18:06:26.51       & -21:35:46.6   & 38.1      \\
 G10.21-0.31        & 18:09:20.63       & -20:15:04.5   & 12.8      \\
 G10.21-0.32        & 18:09:24.52       & -20:15:41.4   & 12.8      \\
 G10.15-0.34        & 18:09:21.38       & -20:19:32.8   & 12.8      \\
 G10.61-0.33        & 18:10:15.62       & -19:54:46.6   & 74.0      \\
 G12.19-0.12        & 18:12:41.67       & -18:25:19.8   & 27.6      \\
 G13.18+0.06        & 18:14:00.92       & -17:28:41.2   & 51.6      \\
 G15.01-0.67        & 18:20:21.22       & -16:12:42.2   & 26.2      \\
 G15.03-0.65        & 18:20:18.80       & -16:11:22.6   & 26.2      \\
 G15.01-0.69        & 18:20:24.22       & -16:13:22.8   & 26.2      \\
 G18.17-0.30        & 18:25:07.53       & -13:14:32.7   & 54.9      \\
 G18.21-0.34        & 18:25:21.55       & -13:13:39.5   & 54.9      \\
 G23.41-0.23        & 18:34:45.74       & -08:34:21.2   & 104.2     \\
 G23.42-0.23        & 18:34:48.16       & -08:33:56.1   & 104.2     \\
 G23.44-0.18        & 18:34:39.25       & -08:31:36.2   & 104.2     \\
 G27.29+0.15        & 18:40:34.70       & -04:57:18.1   & 26.0      \\
 G27.31+0.18        & 18:40:32.45       & -04:55:03.8   & 26.0      \\
 G29.97-0.05        & 18:46:12.25       & -02:39:05.9   & 100     \\
 G34.81-0.28        & 18:53:20.148      & +01:28:31.5   & 59.0      \\
 G35.19-1.73        & 19:01:45.45       & +01:13:21.5   & 42.4      \\
 G81.74+0.59        & 20:39:00.37       & +42:24:36.6   & -3.1      \\
\hline
         &   IRDCs      &      &                                 \\
\hline
 G11.11-0.12 P1 & 18:10:34.04      & -19:21:4    & 29.2      \\
 G11.11-0.12 P3 & 18:10:07.25      & -19:27:2    & 29.2      \\
 G11.11-0.12 P4 & 18:10:07.25      & -19:28:4    & 29.2      \\
 G19.30+0.07 P1 & 18:25:58.14       & -12:04:4    & 26.3      \\
 G19.30+0.07 P2 & 18:25:52.69       & -12:04:4    & 26.3      \\
 G28.34+0.06 P1 & 18:42:50.9        & -04:03:1      & 78.4      \\
 G28.34+0.06 P2 & 18:42:52.4        & -03:59:5      & 78.4      \\
 G33.71-0.01    & 18:52:53.81       & +00:41:06.4    & 104.2     \\
 G79.34+0.33    & 20:32:26.20       & +40:19:40.9    & 0.1       \\

\hline
\end{tabular}
\label{tab:source_list}
\end{table*}

\subsection{Results of the \AMM\ and \DAMM\ single dish observations \label{deutn1}}

The ratio of the brightness temperatures of the \AMM\ (1,1) and (2,2)
transitions, along with the optical depth, can be used to estimate the
rotational temperature. For temperatures $<$20~K, which are typical for
cold dark clouds, the rotational temperature closely follows the gas
kinetic temperature \citep{walmsley1983:nh3}.  Like \AMM, its
isotopologue \DAMM\ also has hyperfines which allows the estimation of
optical depth and hence the column density.  Thus
the fractionation ratio can be estimated assuming that \DAMM\ and
\AMM\ are co-spatial and hence have the same gas temperature.

Out of the 32 sources that were observed, \DAMM\ with hyperfines were detected in 22
sources with S/N ratio $> 5\sigma$ while we have a 100\%
\AMM\ (1,1) and (2,2) detection. While we detected the (3,3) lines in most of the sources, the (4,4) lines were not detected. 
The fits to the hyperfines for \AMM\ and \DAMM\ were
done using CLASS methods \AMM\ (1,1) and HFS respectively
\citep{forveille1989:class}. The line parameters from the resultant
fits are given in Table~\ref{tab:nh3_line parameter} and Table~\ref{tab:nh2d_line parameter}. For those sources
with a high uncertainty in the \DAMM\ optical depth, the main line
is fitted with a single Gaussian and the integrated intensity is
quoted. The spectra are shown in Fig.~\ref{fig:nh2d_nh3}.

 The basic physical parameters, namely the rotational temperature, the
kinetic temperature and ammonia column density were derived using the
standard formulation for \AMM\ spectra \citep{bachiller1997:nh3}. The
expressions used to estimate the column densities from the radiative
transfer equations for all other molecules including \DAMM\ are given
in Appendix A.  The deduced physical parameters are given in
Table~\ref{tab:nh2d_ratio}. The uncertainties given in brackets are
formal errors obtained by a Gaussian error propagation. We have
derived the \DAMM\ column density solely based on the ortho
transition.  We assume that the ortho and para transitions are
in LTE.

For the few cases with significant detection of the para state we find
that ratio of the integrated line intensity (\tdv) is approximately
3 consistent with a thermalised ortho to para ratio of 3.

For the column density determination of \DAMM, we have averaged
over the 9 points of the map to compare to the \AMM\
results. Essentially the resolution of the \DAMM\ observations has
been degraded to that of the \AMM\ observations. A caveat is that a map obtained with 10\arcsec\ spacing at the IRAM 30m does
not completely sample the 40\arcsec\ beam of the Effelsberg
telescope. However, it is very unlikely that there is significant
emission outside the beam of the Effelsberg telescope (i.e. its FWHM) because a)we find that
the brightest peak is always at the center of the map, b) box
averaging delivers a similar result to a 1 D Gaussian
weighted average c) the peak dust emission positions
have been chosen as the centre coordinates for these observations.

 The
rotational temperatures are in a range from 11 to 23~K, with ammonia
column densities from 1 to 5$\times 10^{15}$ \percc.  We obtain
[\DAMM/\AMM] ratios from 0.004 -- 0.66. A deuteration factor of 66\%
in \AMM\ is much larger than the largest reported value so
far,$\sim$33\% by \citet{hatchell2003:nh2d}. In 10 out of the 22
sources with \DAMM\ detection, we obtain abundance ratios $\le$0.02
while the rest of the sources have a high degree of deuteration
($\ge$13\%). These 10 sources with low ratios show optically thin
\DAMM\ emission. 

\begin{figure*}
\centering
 \includegraphics[height=\linewidth,angle=-90]{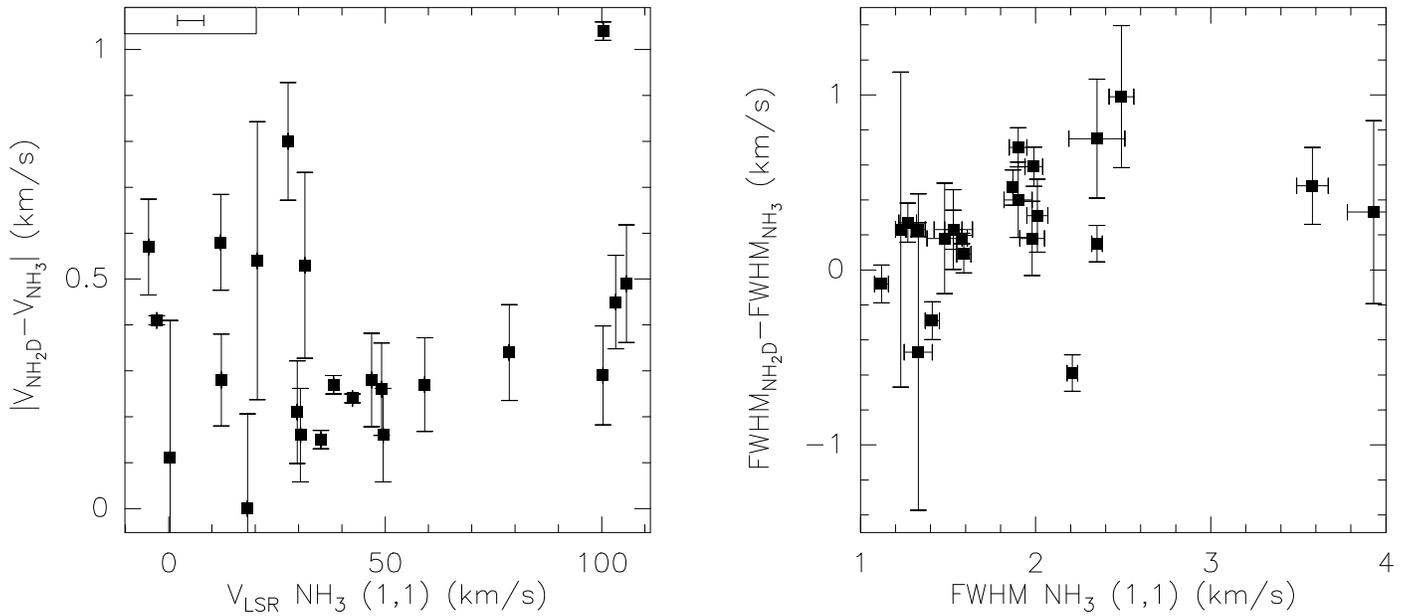}
 \caption{Left panel: Comparison of the LSR velocities of \AMM\ and \DAMM. The median spectral resolution after Hanning smoothing is marked in the upper left corner of the panel. Right panel:  Comparison of \AMM\ and \DAMM\ line widths. Points with no y error bars are those for which the line width has been fixed based on the hfs fit to the optically thin emission, to derive the \DAMM\ integrated intensity.}
\label{fig:nh2d_dv}
\end{figure*}

In Fig.~\ref{fig:nh2d_dv}, the correlation between the velocities and
line widths of the \DAMM\ and \AMM\ emission is shown.  Velocities and
line widths are clearly different in some sources, with the \DAMM\
line widths being smaller in most cases.  This could be either due to
\DAMM\ and \AMM\ tracing different regions or the slightly different
beams at the two frequencies, in which case, clumping might contribute
to a larger line width.  It is interesting to note that the critical
density of the reported \DAMM\ transition is a factor 47 higher than
that of the corresponding \AMM\ transition, assuming similar
collisional rates. However, a more likely explanation is that
\AMM\ could be originating from the more disturbed protocluster
environment, while \DAMM\ might be tracing the gas in precluster cores
(Pillai et al.\ in prep.).

\begin{table*}[h]
\caption{\DAMM\ and \AMM\ column densities, rotational temperature and fractionation }
\vspace{1em}
\begin{center}
\begin{tabular}{lcccc}
\hline
\hline
source   & $T_{\rm rot}$ $^{\rm a}$  & $N_{\rm NH_3}$ $^{\rm b}$                 & $N_{\rm NH_2D}$  $^{\rm c}$                & [\DAMM/\AMM]$^{\rm d}$             \\
         &  K          & $10^{15} (10^{14})~\rm cm^{-2}$ & $10^{13}~\rm cm^{-2}$  &                          \\
\hline
G8.13+0.25    & 18.0    (3.0)   &       1.8     (0.4)  &   1.7   (0.1 ) &         1.0   (0.2 ) \\  
              & 17.4    (2.3)   &       1.6     (0.3)  &   2.4   (0.1 ) &         1.5   (0.3 ) \\ 
G8.68-0.37    & 12.1    (0.8)   &       1.3     (0.1)  &   39.4  (12.4) &         30.3  (9.8 ) \\ 
G8.71-0.37    & 13.3    (1.0)   &       1.5     (0.2)  &   59.7  (10.6) &         39.8  (8.9 ) \\ 
G10.21-0.31   & 15.4    (1.5)   &       2.1     (0.3)  &   70.9  (29.9) &         33.7  (15.0) \\ 
G10.21-0.32   & 14.1    (0.7)   &       2.1     (0.1)  &   51.3  (22.3) &         24.4  (10.7) \\ 
G10.61-0.33   & 17.8    (1.5)   &       2.3     (0.2)  &   1.0   (0.1 ) &         0.4   (0.1 ) \\ 
G11.11-0.12P1 & 12.5    (1.2)   &       1.4     (0.2)  &   54.0  (23.5) &         38.6  (17.7) \\ 
G11.11-0.12P4 & 10.6    (1.6)   &       1.8     (0.5)  &   1.7   (0.3 ) &         0.9   (0.3 ) \\ 
G12.19-0.12   & 12.6    (0.3)   &       4.4     (0.2)  &   5.6   (0.2 ) &         1.3   (0.1 ) \\ 
G13.18+0.06   & 17.4    (1.3)   &       2.8     (0.2)  &   88.9  (23.2) &         31.7  (8.6 ) \\ 
G18.17-0.30   & 15.8    (1.2)   &       1.8     (0.2)  &   119.7 (31.5) &         66.5  (19.0) \\ 
G18.21-0.34   & 15.5    (1.2)   &       1.7     (0.2)  &   47.3  (19.7) &         27.8  (12.0) \\ 
G23.41-0.23   & 18.8    (1.8)   &       2.5     (0.3)  &   77.7  (51.3) &         31.1  (20.8) \\ 
G23.44-0.18   & 23.3    (1.5)   &       5.2     (0.4)  &   5.0   (0.2 ) &         1.0   (0.1 ) \\ 
G27.29+0.15   & 17.8    (1.7)   &       3.0     (0.3)  &   1.4   (0.2 ) &         0.5   (0.1 ) \\ 
G28.34+0.06P2 & 15.7    (2.0)   &       2.2     (0.3)  &   4.1   (0.2 ) &         1.9   (0.3 ) \\ 
G29.97-0.05   & 13.7    (1.4)   &       2.2     (0.3)  &   67.7  (15.2) &         30.8  (8.1 ) \\ 
G33.71-0.01   & 19.5    (4.6)   &       3.0     (0.8)  &   1.4   (0.1 ) &         0.5   (0.1 ) \\ 
G34.81-0.28   & 16.6    (1.0)   &       1.6     (0.1)  &   2.7   (0.2 ) &         1.7   (0.2 ) \\ 
G35.19-1.73   & 17.0    (0.9)   &       1.8     (0.1)  &   74.0  (11.0) &         41.1  (6.5 ) \\ 
G79.34+0.33   & 13.7    (1.0)   &       1.4     (0.1)  &   17.9  (15.9) &         12.8  (11.4) \\ 
G81.74+0.59   & 18.4    (1.1)   &       2.7     (0.2)  &   92.9  (17.6) &         34.4  (7.0 ) \\ 
\hline
\end{tabular}
\end{center}
$^{\rm a}$ Rotational temperature derived from \AMM\ observations. \\
$^{\rm b}$ \AMM\ column density. \\
$^{\rm c}$ \DAMM\ column density. \\
$^{\rm d}$ Ratio of \AMM\ and \DAMM\ column densities in percentage. \\
Note: Similar notations are used for other molecules elsewhere in this paper.
\label{tab:nh2d_ratio}
\end{table*}

\subsection{Results of the CO observations \label{res:depn1}}


The \CO\ (1 $\rightarrow$ 0) and (2 $\rightarrow$ 1) observations were
performed simultaneously with \DAMM\ ($1_{1,1}-1_{0,1}$). Figure \ref{fig:c18o_spec} shows the spectra toward all
sources. Note that there are secondary features in the \CO\ spectra
for a few sources, likely to be line of sight components not seen in
dense gas like \AMM. In Table~\ref{tab:c18o_line_parameter} we list
the observed line parameters based on Gaussian fits, and, in
Table~\ref{tab:depl}, the column density and excitation temperature
estimates (see Appendix A) and the \CO\ abundances. The \CO\ excitation temperature has been derived from the \CO\ $J=1$--0 and $J=2$--1 line ratios. 

Based on the dust continuum and CO observations for the observed sample, the depletion can be studied for the first time on a sample of massive clumps.
Recent observational studies on the condensation of CO in low mass prestellar
cores reveal that in the initial cold and dense evolutionary stage, CO
is heavily depleted onto the dust grains (\citealt{bacmann2002:depln}, \citealt{savva2003:co:depln}). The
depletion factor $\eta$ is defined as
\begin{equation}
 \eta  \cdot {\frac{N_{\rm C^{18}O}}{[N_{\rm H_2}]}}_{\rm observed}=
{\frac{N_{\rm C^{18}O}}{[N_{\rm H_2}]}}_{\rm canonical}
\label{eq:depln_fac}
\end{equation}
The $\rm{C^{18}O}$ column density, ${N_{\rm C^{18}O}}$, is derived from the observed
integrated intensity (see Appendix~A) and assumimg that the gas temperature is equal to the excitation temperature derived from the \CO\ line ratios.

The effective ${\rm H_2}$ column density \citep{launhardt1996:thesis} is
calculated from
\begin{equation} N({\rm H_2}) =
{\frac{6.2\times10^{16}S_{\nu}\lambda^{3}\ \rm
e^{\frac{1.44\times10^{4}}{T_d\lambda}}}{\kappa_{\rm
m}(\lambda)\theta^{2}}}{\frac{Z{\odot}}{\rm Z}}
\label{eq:h2}
\end{equation} where $S_{\nu}$ is the flux density in
Jy/beam. $\lambda$ the wavelength in $\mu$m, $\rm Z/Z{\odot}$ is the
metalicity relative to the solar metalicity(we assume 1) and $\theta$
is the FWHM of the beam at wavelength $\lambda$.  We used an opacity
$\kappa_{\rm m}$ of 1.85 $\rm cm^2/g$ of dust at $850 ~\mu$m for dust
grains with thick ice mantles and gas density $\rm n(H)= 10^6~$\percc
\citep{ossenkopf1994:opacities}. The column density was derived
after smoothing the 850 $\mu$m dust continuum emission to the 20$''$
resolution of the \CO\ data (Thompson et al.\ 2007 in prep).

\begin{table*}[p{2.1cm}]
\begin{center}
\caption{\CO\ column density and abundance from ($J=1$--0) and ($J=2$--1) transitions }
\vspace{1em}
\begin{tabular}{|l|c|c|c|c|c|}
\hline
Source  &  \multicolumn{1}{c|}{$T_{\rm ex}$} &\multicolumn{1}{c|}{$N_{\rm C^{18}O}$} & \multicolumn{1}{c|}{N(${\rm H_2}$)$^{\rm a}$} & \multicolumn{1}{c|}{$\chi_{\rm C^{18}O}$ $^{\rm b}$} & \multicolumn{1}{c|}{$\eta$$^{\rm c}$}   \\
\hline
        &  \multicolumn{1}{c|}{K} &\multicolumn{1}{c|}{$10^{16} (10^{15})~\rm cm^{-2}$} & \multicolumn{1}{c|}{$10^{23}$~cm$^{-2}$} & \multicolumn{1}{c|}{10$^{-8}$} & \multicolumn{1}{c|}{}  \\
\hline
G8.13+0.25     &  11.8    & 1.6     (0.1)     &  2.6     & 6.2     & 2.7 \\
G8.68-0.37     &  8.5     & 2.4     (0.9)     &  7.9     & 3.0     & 5.6 \\
G8.71-0.37     &  4.2     & 1.8     (0.5)     &  3.5     & 5.2     & 3.2 \\
G10.15-0.34    &  19.1    & 2.1     (0.4)     &  3.8     & 5.4     & 3.1 \\
G10.21-0.31    &  8.4     & 1.4     (0.2)     &  2.2     & 6.4     & 2.7 \\
G10.21-0.32    &  8.0     & 2.6     (0.3)     &  7.4     & 3.6     & 4.8 \\
G10.61-0.33    &  9.1     & 1.0     (0.3)     &  2.7     & 3.8     & 4.5 \\
G11.11-0.12P1  &  6.4     & 0.6     (0.1)     &  1.1     & 5.5     & 3.1 \\
G11.11-0.12P3  &  6.9     & 0.5     (0.3)     &  0.8     & 6.2     & 2.7 \\
G11.11-0.12P4  &  6.6     & 0.5     (0.2)     &  1.4     & 3.5     & 4.8 \\
G12.19-0.12    &  11.6    & 1.2     (0.2)     &  2.4     & 5.1     & 3.3 \\
G13.18+0.06    &  15.7    & 1.2     (0.6)     &  6.3     & 1.8     & 9.3 \\
G15.01-0.67    &  32.5    & 6.0     (0.4)     &  16.6    & 3.6     & 4.7 \\
G15.01-0.69    &  52.2    & 4.5     (0.6)     &  11.6    & 3.9     & 4.4 \\
G15.03-0.65    &  51.7    & 5.5     (0.6)     &  14.2    & 3.9     & 4.4 \\
G18.17-0.30    &  14.3    & 1.5     (0.8)     &  3.3     & 4.6     & 3.7 \\
G18.21-0.34    &  10.9    & 0.7     (0.4)     &  2.5     & 3.0     & 5.8 \\
G23.41-0.23    &  8.4     & 3.4     (0.8)     &  9.3     & 3.6     & 4.7 \\
G23.42-0.23    &  9.0     & 3.1     (0.3)     &  3.8     & 8.0     & 2.1 \\
G23.44-0.18    &  9.2     & 3.0     (0.5)     &  7.7     & 3.9     & 4.3 \\
G27.29+0.15    &  8.8     & 1.2     (0.2)     &  1.7     & 6.9     & 2.5 \\
G27.31+0.18    &  8.3     & 0.5     (0.2)     &  1.7     & 2.7     & 6.3 \\
G28.34+0.06P1  &  4.8     & 1.3     (0.5)     &  1.5     & 8.6     & 2.0 \\
G28.34+0.06P2  &  8.7     & 1.4     (0.5)     &  7.8     & 1.8     & 9.6 \\
G29.97-0.05    &  9.3     & 1.3     (0.7)     &  2.6     & 4.8     & 3.5 \\
G79.34+0.33    &  8.4     & 0.7     (0.1)     &  2.1     & 3.5     & 4.9 \\
G81.74+0.59    &  16.0    & 1.8     (0.2)     &  4.7     & 3.8     & 4.4 \\
\hline
\end{tabular}
\label{tab:depl}
\end{center}
$^{\rm a}$ ${\rm H_2}$ column density derived from the 850 $\mu$m dust
continuum flux measured using SCUBA.\\ The dust temperature is assumed
to be equal to the gas temperature derived from \AMM\ (1,1) and (2,2)
measurements as given in Table~\ref{tab:nh2d_ratio}. \\ $^{\rm b}$
abundance of \CO\ relative to H$_2$.\\ $^{\rm c}$ depletion factor
compared to the canonical value of 1.7$\cdot 10^{-7}$
\citep{frerking1982:co}.\\
\end{table*}

\begin{figure}
\centering
 \includegraphics[height=0.7\linewidth,angle=-90]{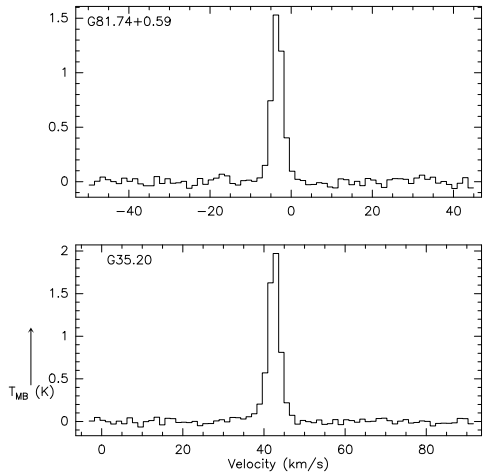}
 \caption{\COs\ ($J=2$--1) spectra toward G35.19-1.73 and G81.74+0.59 observed with the IRAM 30m telescope.}
\label{fig:c17o_spec}
\end{figure}

In these dense clumps it is possible
that the \CO\ line may be optically thick and hence derived column densities could be a lower limit.
In order to investigate this possibility, we observed the rarer
isotopomer \COs\ toward two of our sources, G35.19-1.73 and G81.74+0.59 as
shown in Fig.~\ref{fig:c17o_spec}. The line parameters are given in
Table~\ref{tab:c17o_line_parameter}. In Sect.~\ref{subsec:co_thin}, we discuss how \COs\ is used as a more reliable tracer 
of the optical depth.

\begin{table}[p{0.1cm}]
\caption{\COs\ ($J=2$--1) line parameters }
\small
\begin{tabular}{lcc}
\hline
 Source       &  \tdv\           & $\Delta$v          \\  
              &   K~\kms\         & \kms\              \\    
\hline

G35.19-1.73 & 7.1 (0.1) & 3.3 (0.1)      \\
G81.74+0.59 & 5.5 (0.1) & 3.3 (0.1)     \\

\hline
\end{tabular}
\label{tab:c17o_line_parameter}
\end{table}

\subsection{Results of the \htcn\ and \hcfn\ observations \label{res:hcn}}
We also observed the isotopomeric species of HCN, \htcn\ and \hcfn\ in
our 3~mm setup with the IRAM 30m telescope.  \htcn\ was detected in
almost all sources except G19.30P1/P2 and G10.15-0.34. In
Fig.~\ref{fig:h13cn_hc15n}, we show the spectra toward all sources
where both \htcn\ and \hcfn\ are detected with sufficient S/N ($\ge
3~\sigma$). The \htcn\ line has hyperfines, which allows to estimate the
optical depth. However, we find that \htcn\ is optically thin in
almost all cases. In two sources with moderate optical depths, the
uncertainty in $\tau$ is too large to derive meaningful constraints on the
column density. The line parameters are given in
Table~\ref{tab:h13cn_line parameter} along with their column
densities. The column density is computed using the gas temperature
derived from \AMM\ as the excitation temperature (see Appendix A).

\begin{table*}[h]
\begin{center}
\caption{\htcn\ and \hcfn\ line parameters and column density}
\vspace{1em}
\begin{tabular}{l|cccc|ccc|}
\hline
              &   \multicolumn{4}{c|}{\htcn\  $J=1$--0, $F=2$--1} &  \multicolumn{3}{c|}{\hcfn\   $J=1$--0}  \\
 Source  & \multicolumn{1}{c}{\tdv$^{\rm a}$} & \multicolumn{1}{c}{$\Delta$v} & \multicolumn{1}{c}{$T_{\rm rot}$} & \multicolumn{1}{c|}{$N_{\rm H^{13}CN}$} & \multicolumn{1}{c}{\tdv} & \multicolumn{1}{c}{$\Delta$v} & \multicolumn{1}{c|}{$N_{\rm HC^{15}N}$} \\
\hline
         & \multicolumn{1}{c}{K~\kms} & \multicolumn{1}{c}{\kms} & \multicolumn{1}{c}{K} & \multicolumn{1}{c|}{$10^{12}~\rm cm^{-2}$} & \multicolumn{1}{c}{K~\kms} & \multicolumn{1}{c}{\kms} & \multicolumn{1}{c|}{$10^{12}~\rm cm^{-2}$} \\
\hline
G10.21-0.31     &   0.9 (0.1)      & 3.5      & 15.4  (1.5)   &  12.2   (1.1) & 0.2     ($<$0.1)  & 1.2     (0.4)     &  0.5    (0.1)     \\
G10.21-0.32     &   0.7 (0.1)      & 2.4      & 14.5  (1.7)   &  9.3    (1.0) & 0.6     (0.2)     & 5.6     (2.2)     &  1.3    (0.6)     \\
G13.18+0.06     &   3.0 (0.1)      & 3.7      & 17.4  (1.3)   &  42.7   (2.5) & 0.7     (0.1)     & 3.3     (0.4)     &  1.9    (0.2)     \\
G15.01-0.67     &   7.3 (0.2)      & 3.4      & 24.3  (1.5)   &  129.5  (7.1) & 2.6     (0.2)     & 2.8     (0.3)     &  8.6    (0.9)     \\
G15.01-0.69     &   4.8 (0.2)      & 2.7      & 24.3  (1.3)   &  85.2   (4.3) & 1.6     (0.1)     & 2.4     (0.2)     &  5.3    (0.4)     \\
G15.03-0.65     &   3.6 (0.1)      & 3.6      & 20.4  (2.4)   &  55.9   (4.9) & 1.3     (0.1)     & 3.6     (0.0)     &  4.4    (1.0)     \\
                &   1.6 (0.1)      & 1.7      & 16.6  (2.9)   &  21.6   (2.6) & 0.3     (0.1)     & 1.7     (0.0)     &  0.9    (0.2)     \\ 
G23.41-0.23     &   1.3 (0.1)      & 2.0      & 18.8  (1.8)   &  18.6   (1.4) & 0.5     (0.1)     & 2.6     (0.5)     &  1.3    (0.2)     \\
G23.42-0.23     &   1.4 (0.1)      & 3.4      & 19.0  (2.0)   &  20.3   (1.7) & 0.4     (0.1)     & 2.6     (0.4)     &  1.1    (0.2)     \\
G28.34+0.06P2   &   1.0 (0.1)      & 2.4      & 15.7  (2.0)   &  13.8   (1.2) & 0.7     (0.1)     & 3.7     (0.5)     &  1.8    (0.2)     \\
G35.19-1.73     &   1.9 (0.1)      & 2.4      & 17.0  (0.9)   &  27.1   (1.3) & 0.6     (0.1)     & 2.0     (0.3)     &  1.4    (0.2)     \\
G79.34+0.33     &   0.8 ($<$0.1)   & 1.5      & 13.7  (1.0)   &  10.0   (0.5) & 0.2     ($<$0.1)  & 1.2     (0.2)     &  0.5    (0.1)     \\
G81.74+0.59     &   3.2 (0.1)      & 2.8      & 18.4  (1.1)   &  46.7   (2.3) & 1.1     (0.1)     & 2.8     (0.2)     &  3.0    (0.2)     \\

\hline
\end{tabular}
\label{tab:h13cn_line parameter}
\end{center}
$^{\rm a}$ Integrated line intensity for the \htcn\ main group alone. \\ 
\end{table*}

\section{Analysis and Discussion \label{sec:analysis}}

\subsection{CO Depletion \label{sec:depln_analysis}}

The derived
\CO\ abundances are a factor of 2 --12 lower than the canonical value of 1.7$\cdot 10^{-7}$
\citep{frerking1982:co}. \citet{wu2005_irdcg79} find similar depletion factors for 
the infrared dark cloud G79.2+0.38, another of the clumps in the cloud in the present study.

\subsubsection{Is \CO\ optically thin? \label{subsec:co_thin}}

As mentioned in Sect.~\ref{res:depn1}, the CO column density as
determined from \CO\ transitions might be underestimated if \CO\
is optically thick. However, the rarer isotopomer \COs\ is expected to
be optically thin.  The canonical value of the relative abundance of \CO\
w.r.t \COs, $A(18,17)$, is 3.65 \citep{wilson1994:araa} in the ISM.  As
discussed in \citet{kramer1999:ic5146}, the measured ratios of the
integrated intensities of the \CO\ and \COs\ lines can then be used to
determine the \CO\ optical depth. 

Due to the limited observing time, we observed the \COs\ line only in the two brightest \CO\ sources.
If $R^{18,17}$ and $\tau^{18,17}_{18}$ denotes the ratio of
the integrated intensities of the \CO\ and \COs\ lines and the optical depth of \CO\ as derived from
the ratio, then
\begin{equation}
R^{18,17} =
\frac{1-\exp(-\tau^{18,17}_{18})}{1-\exp(-\tau^{18,17}_{18}/{A(18,17)})}
\, .
\label{eq:c17o_c18o}
\end{equation}

The assumptions involved in deriving Eq.\ref{eq:c17o_c18o}, are that
both isotopomers have the same excitation temperatures and beam
filling factors.

For G35.19-1.73, we find that $\tau^{18,17}_{18}=1.35$, which implies
that the correction factor for the column density given by
$\tau/(1-\exp(-\tau))=1.8$, is clearly smaller than the observed
depletion. However, for G81.74+0.59, the ratio itself $R^{18,17} \sim
6$, suggesting that Eq.(\ref{eq:c17o_c18o}) is not valid anymore. It
is unlikely to be due to different excitation conditions or the extent of
emission of \CO\ and \COs. It is indeed possible that there are real variations in
the ratio of $^{18}\rm O$ to $^{17}\rm O$, a possibility which needs to be investigated.
Additionally, the observed relatively small observed
peak brightness temperatures for \CO\ exclude high optical depth in this line.
Therefore, to the extent that we assume G35.19-1.73 to be representative of our
sample, we may conclude that \CO\ optically depth is ``moderate''.

\subsection{\AMM\ deuteration}

How do we account for the high deuteration we observe in these
pre/protostellar clumps? There are two main pathways to bring about
deuteration (\citealt{rodgers2001:nhd2},
\citealt{millar2002:modelling_deutn}, \citealt{millar2003:deutn},
\citealt{roueff2005:deutn}) ; a) gas-phase reactions, b) production of
deuterium bearing molecules on grain surfaces. Accretion of
neutrals onto the dust grains that would otherwise destroy ${\rm H_2
D^+}$ enhances deuteration. Up to now all these processes have been used
to explain fractionation in different sources.

In pre/protostellar cores, ion-molecule exchange reactions prevalent
at low temperatures coupled with the depletion of CO from gas phase is
preferred over grain production of highly deuterated species at
temperatures of 20~K
(\citealt{shah2001:nh2d},\citealt{caselli2003:h2d+} ). 
The freeze-out of molecules from the gas onto the dust grains,
particularly that of heavy molecules like CO in the cold initial
phase has been predicted by chemical models
(\citealt{flower2005_freeze_out}, \citealt{rodgers2001:nhd2}, \citealt{roberts2000a,roberts2000b}, \citealt{brown1989a:deutn}, \citealt{watson1976:ism_reactions}). These predictions have been validated by observations of molecular freeze out onto dust grains in low-mass starless dense cores (\citealt{caselli1999:depln}, \citealt{tafalla2002:depletion}). The primary fractionation reaction that dominates at
low temperatures ($< 20$~K) is 
\begin{equation}
\rm{H_3^+ + HD \rightarrow H_2 D^+ + H_2 + \Delta E_1}
\label{eq:h2d+}
\end{equation}
Neutral molecules like CO can destroy ${\rm H_2 D^+}$,
thereby lowering the deuterium enhancement. Therefore, the depletion
of CO, the second most abundant molecule, from the gas phase can lead
to an enhancement in the $[{\rm H_2 D^+}]/[{\rm H_3^+}]$ ratio and
thereby the molecular D/H ratios.

\begin{figure}
\centering
\includegraphics[height=\linewidth,angle=-90]{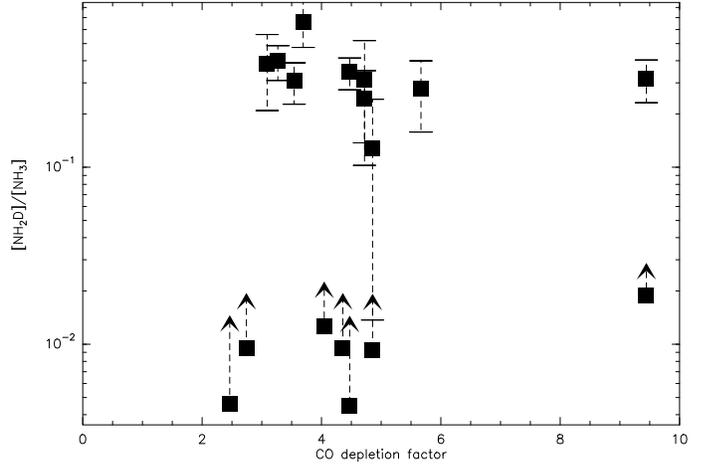}
 \caption{Comparison of \AMM\ fractionation and CO depletion factor for the dense clumps in our sample. The error bar on CO depletion factor is quite large mainly due to the uncertain dust opacities adopted (see Sect.~\ref{sec:depln_analysis}).}
\label{fig:depln}
\end{figure}

\citet{roberts2000b} show that at 10~K, accretion of neutrals onto the dust grains,
especially CO, leads to the formation of doubly deuterated molecules
such as $\rm NHD_2$ and $\rm{D}_2{\rm CO}$. Based on these arguments,
we expect to see a correlation between deuteration and CO depletion.
In Fig.~\ref{fig:depln}, we compare the \AMM\ fractionation and the
degree of CO depletion (depletion factor).  As seen from the
correlation plot, there is a large scatter in the values and we do not
see any obvious trend of high depletion and deuteration.
Nevertheless, the main result is that we find very high deuterium
fractionation and CO depletion. 

Depletion could well occur on scales much smaller than the beam for
sources at the distances typical for these sources(several kpc). In
that case, the CO depletion that is measured as an average over the telescope beam is an underestimate of the true
depletion. The \DAMM\ emission, on the other hand, is probably
dominated by the dense clumps.  This could explain why the CO
depletion does not appear to track the deuteration in
Fig.~\ref{fig:depln}. The uncertainties on the CO depletion factors are
quite large, mainly due to the uncertainty in the $\rm H_2$ column
density estimate from the 850$\mu$m dust continuum. The largest
uncertainty is a factor of 4 in the dust opacity at 850$\mu$m (\citealt{ossenkopf1994:opacities}, \citealt{kruegel1994},
\citealt{draine1984}). But we note that with the choice of the
Ossenkopf \& Henning opacities, the derived column densities are
already lower than for e.g. \citet{savage1979} dust properties. The
two studies other than ours that directly compare CO depletion and
deuterium fraction are by \citet{bacmann2003:depln} and
\citet{crapsi2005:deutn_depn}. Bacmann et al. use the ${\rm D_2CO}$ to
${\rm H_2CO}$ ratios to determine the deuteration. They claim to
find a correlation between ${\rm D_2CO}$ to ${\rm H_2CO}$ ratios and
depletion and argue that the presence of a significant amount of O
(which is also an important ${\rm H_3^+}$ destroyer) in the gas phase
still undepleted might be responsible for the large scatter
observed. Crapsi et al. observe N$_2$H$^+$ and N$_2$D$^+$ toward 31
low-mass starless cores and find a good correlation between deuterium
fractionation and CO depletion.

If $f(\rm X)$ is the fractional abundance of species X relative to $\rm H_2$ and
$R({\rm XD})$ is the abundance ratio of ${\rm XD}$ relative to ${\rm XH}$,
then under the assumption of steady-state and equating formation and destruction in
Eq.(\ref{eq:h2d+}), one finds \citep{millar2003:deutn}
\begin{equation}
R({\rm H_2 D^+})= S_{\rm H_2 D^+}(T) f(HD)
\, .
\label{eq:h2d+_t1}
\end{equation}

where $S_{\rm H_2 D^+}$(T) is a function of the different rate coefficients.
(forward, backward and dissociative
recombination rates) given by 
\begin{equation}
S_{\rm H_2 D^+}(T)= \frac{k_{1f}}{k_{1r}+\alpha_1 f({\rm e})+\Sigma k_{\rm M}f({\rm M})} 
\, .
\label{eq:h2d+_t2}
\end{equation}

Here, $k_{1f}$, $k_{1r}$ are the forward and reverse rate coefficients
such that $k_{1r}=k_{1f}\exp(-\Delta E_1/{\rm T})$ and $\alpha_1
\propto 1/\sqrt{T}$ is the dissociative recombination rate, $f({\rm
e})$ is the electron fraction and $\Sigma k_{\rm M}$ is the rate
coefficient of ${\rm H_2 D^+}$ with species M. $\Delta E_1$ is the
exothermal energy ($\sim 220$~K), also given in Eq.(\ref{eq:h2d+}).

\begin{figure}
\centering
 \includegraphics[height=\linewidth,angle=-90]{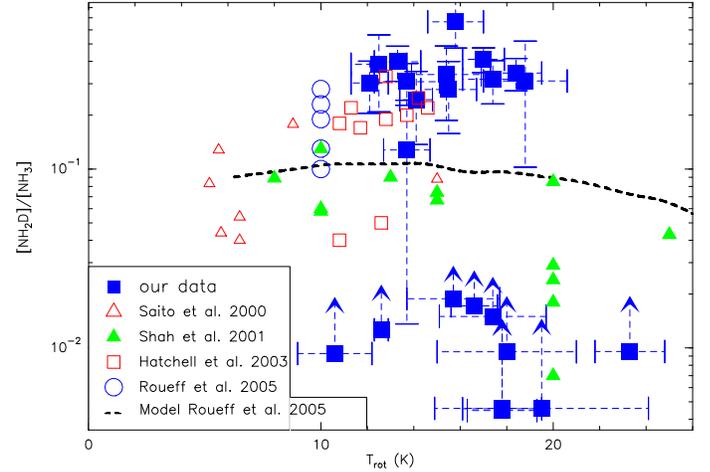}
 \caption{Plot of \AMM\ fractionation versus gas temperature as
 derived from \AMM. The dashed line is the latest gas phase model
 predictions \citep{roueff2005:deutn}. The filled and unfilled squares
 mark the SCAMPS sources and \citet{hatchell2003:nh2d} respectively;
 filled and unfilled triangles, the values found toward low mass
 pre/protostellar cores by \citet{shah2001:nh2d} and
 \citet{saito2000:nh2d} respectively and the unfilled circles mark
 \citet{roueff2005:deutn} sources.}

\label{fig:deutn}
\end{figure}

At low temperatures ($\rm T < 20$~K) observed in cold cores, the
primary fractionation reaction is dominated by the forward reaction
in Eq.(\ref{eq:h2d+}). At higher temperatures, the reverse reaction
becomes important in removing ${\rm H_2 D^+}$ very rapidly from the
gas phase. Consequently, the decrease in primary fractionation
produces a corresponding decrease in secondary fractionation reactions
responsible for deuterium enhancements in molecules. Therefore, one naively
expects to find an anti-correlation between temperature and deuterium
fractionation. In Fig.~\ref{fig:deutn}, we plot the \AMM\ fractionation
against the temperature for the high mass candidate pre/protostellar clumps (this paper) and 
 for pre/protostellar clumps in the lower mass regime from the literature. 
The latest gas-phase predictions for [\DAMM]/[\AMM]
\citep{roueff2005:deutn} are also plotted. This model assumes a density
of ${\rm n(H_2) = 10^5}$~\percc, carbon and oxygen depletion factors
of 5 and 15 respectively while Nitrogen is kept constant. Our data points nicely
fill in the missing observed data points between 10 --
20~K. Clearly the agreement with the model is moderate. 

The levels of deuteration do fall off rapidly with increasing
temperature above 25~K \citep{roueff2005:deutn}, but at the low
temperatures shown here the dependence is small. The Roueff et
al. models employ fixed H$_2$ densities of around 10$^5$~\percc, which
would be typical averages over the 40\arcsec\ beam for our sources.
However the \DAMM\ emission will be enhanced at the peaks, where
higher densities and higher depletion would be expected to produce
higher levels of deuteration. Depletion, which increases over time
roughly with the increase in density, has a strong influence on
deuteration.  The final levels of depletion and deuteration depend on
the accretion history (eg. \citealt{flower2005_freeze_out}), but it is
clear that higher densities or longer timescales result in higher
levels of both, eg. \citet{roberts2003_enhanced_deutn} derive
[\DAMM]/[\AMM] ratios of 0.4--0.8 for densities of
3$\times10^6$~\percc.

However, most of the low [\DAMM]/[\AMM] ratios and interestingly the
high temperatures are from the sources where \DAMM\ column
densities were estimated from the integrated intensity of the line
and therefore are really lower limits on the [\DAMM]/[\AMM]
ratios. The exact temperature which we measure for a clump of course
depends on the combination of hot/cold gas in the beam, which can
explain the scatter between 15 -- 18~K where in some cases we are
still seeing a lot of deuteration and in others it has already
diminished.

On average, the fractionation observed in sources in our sample is
very high. However, note that few sources at temperatures above 15~K
show an order of magnitude lower deuteration than the model
predictions, while others show extremely high deuterations, much
higher than those found in low mass pre/protostellar cores.

There could be three main reasons for the very low fractionation
estimated for a few sources. First, one of the main assumption in
deriving the [\AMM]/[\DAMM] ratio is that the filling factors for both
molecular transitions are the same.  Any deviation from this
assumption might result in a discrepant ratio.  Second, these sources
might be relatively more evolved. In such a case, the kinetic
temperatures derived from the \AMM\ (1,1) and (2,2) lines might only
be a lower limit.  In such cases where the gas temperature is
roughly higher than 20 K, the rotational temperature does not
represent the real gas temperature. Hence the column density
(deuteration) determined may be under estimated. Third, the sources could be chemically young, hence
the timescale to reach the high deuteration in these sources is larger
than their age. \citet{tafalla2004:young_core} recently
discussed such a chemically young low mass core. It is also possible
that \DAMM\ might be tracing different regions from \AMM\, and that
the temperature derived from \AMM\ may not be the temperature of
\DAMM.

\subsubsection{Upper limits from \DDAMM}

A simple model for grain surface formation of multiply deuterated
molecules is considered by \citet{brown1989b} and predicts the
abundances of deuterated species, in order to differentiate between
the gas and grain chemistry.  According to the Brown and Millar model prediction, the
abundances of deuterated species scale as
[\DDAMM]/[\AMM]$ = \frac{1}{3}$$([\mathrm{NH_{2}D]/[NH_3]})^2$ for grain
surface formation.

Alternatively, \citet{roberts2000a} propose a gas-phase chemistry where 
the effects of the freeze out of gas phase species onto grains is included and
find an enhancement in the fractionation of both singly and doubly deuterated species.

Very deep observations of the \DDAMM\ 1$_{10}$--1$_{01}$ line for two
sources G35.19-1.73 and G81.74+0.59, did not result in any detection at an rms
of $\approx$ 20~mK. Based on the 3~$\sigma$ upper limits, we then derive
[\DDAMM]/[\AMM] (see Appendix A) and find that\\
$([\mathrm{NHD_{2}]/[NH_3])_{observed}<(0.02)_{G35.19}; (0.02)_{G81.74}} $\\
${\rm ({\frac{1}{3} ([\mathrm{NH_{2}D]/[NH_3]})^2)}_{observed}<(0.06)_{G35.19}; (0.04)_{G81.74}}$
$([\mathrm{NHD_{2}]/[NH_2D])_{observed}<(0.05)_{G35.19}; (0.05)_{G81.74}}$\\.

The grain-chemistry model prediction is based purely on
probability arguments assuming that the fractionation is proportional
to gas-phase atomic D/H ratio and that there are three different
pathways for inserting the D atom into the chain of reactions that form
multiply deuterated \AMM. In spite of the simplicity of
the model, the prediction is close to the derived upper limits in both
sources.

\begin{figure}
\centering
 \includegraphics[height=\linewidth,angle=-90]{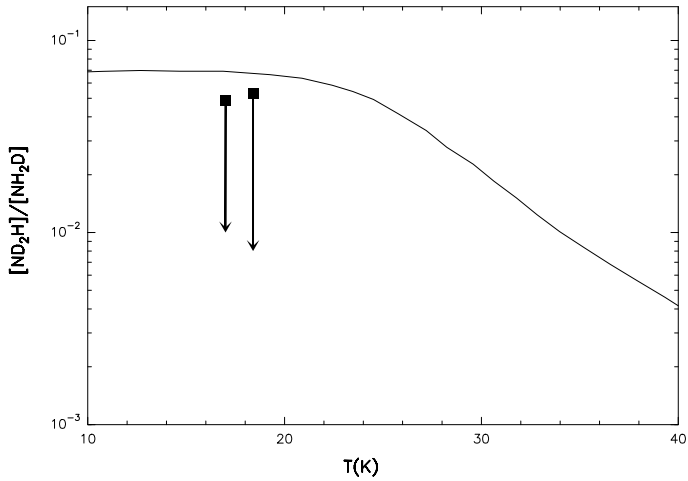}
 \caption{Comparison of obtained upper limits for G35.19-1.73 and G81.74+0.59 with the gas phase model
 predictions (\citealt{roueff2005:deutn}, their model 2).}

\label{fig:nhd2}
\end{figure}

How do the obtained upper limits compare with the predictions of
recent gas-phase deuterium chemistry models?  In Fig.~\ref{fig:nhd2},
we compare the results of model calculations with our observed upper
limits for the ratio of \DDAMM\ to \DAMM\ \citep{roueff2005:deutn}. We
have used their ``model 2'' which gives the highest molecular
fractional abundances for the N-bearing species. In this context it is
also interesting to note that while Roueff et al.\ find that their
model underpredicts the \DDAMM\ to \DAMM\ ratio in low mass prestellar
cores (LDN 134N, LDN 1689N, Barnard 1, LDN 1544), it agrees well with
the observed ratio in our sample.

Hence, we cannot reject active grain or gas-phase chemistry on the
basis of the observed \DDAMM\ upper limits. While a gas-phase model
along with condensation of neutral species onto grain mantles
adequately explains moderate deuteration (\DAMM/\AMM) observed towards
some sources, the very high fractionation we derive remains
unexplained. Although we do not understand the grain-surface
chemistry, particularly in this case involving deuterated ammonia, a
naive extrapolation from the surface-chemistry involving H, D, C and O
and a high grain-surface D/H ratio might explain the significant
enhancement of deuterated ammonia \citep{caselli2004:depletion}
However, the efficacy of this model is strongly dependent on the
gas-phase atomic D/H ratio, which in the extreme case of our sample
implies an enhancement of $\approx 10^{4}$ over the cosmic D/H ratio.

\subsubsection{Comparison with deuteration in other star forming regions}

Although the exact evolutionary status of the sources
(i.e. pre/protostellar) in our sample cannot be determined with
certainty, there are several observational evidences that substantiate
their high mass pre/protostellar nature. The large line broadening ($1<\Delta
v<5$~\kms) observed in all molecular tracers, together with the high
H$_2$ column density ($N({\rm H_2})>10^{23}$~cm$^{-2}$) is
characteristic of massive star forming regions. These sources have masses
between a few hundred to a few thousand solar masses. Their
source-averaged H$_2$ densities are a few 10$^5$~\percc - unlikely for
low-mass objects.  Moreover, the physical properties of these cores
are similar to those associated with UCH{\sc ii} regions
(e.g. \citealt{thompson2006a:scuba_uchii}). Therefore we expect them
to be forming (or on the verge of forming) high-mass stars. The exact
nature of these sources will be discussed in a future paper (Hatchell
et al. in prep.).

The deuteration of a few tens of percent in ammonia is comparable to
the highest which have been measured in the interstellar medium, which
occur in low-mass prestellar cores.  The related ratio
[N$_2$D$^{+}$]/[N$_2$H$^+$] exceeds 20\% in a few prestellar cores
including L1544 \citep{crapsi2004:n2h+_d+_in_l1521f}.  High deuterated ammonia abundances
have also been measured in early-stage protostars: the
[NH$_2$D]/[NH$_3$] ratio reaches over 30\% in protostars in Perseus
\citep{hatchell2003:nh2d}, and detections of triply deuterated ammonia
\citep{vandertak02,lis02} and methanol \citep{parise04} have all been in the
environments of protostars, though deuteration of $\sim 5\%$
is more common \citep{hatchell2003:nh2d,saito2000:nh2d,shah2001:nh2d}. The high
deuteration of our sample supports the idea that these
are the high-mass equivalent of prestellar or protostellar cores.

Hot molecular cores, believed to be the early protostellar stage of
high mass star formation and potentially a later evolutionary stage of
our sources, have a low molecular deuteration fractionation $\sim
10^{-3}$, consistent with ice formation at a higher temperature of
60 -- 80~K and little gas-phase deuteration, as expected at high
temperatures \citep{hatchell98,hatchell99,roberts02}.  High
[NH$_2$D]/[NH$_3$] ratios are observed in the Orion region but only in
the compact ridge and not the hot core \citep{turner90}.  If our
candidate precluster cores are precursors of hot cores, then the
chemistry must radically alter during the later evolution of the hot
core, perhaps with continuing accretion at higher temperatures during
the early protostellar phase dominating the ice production.

\subsection{Variations in HCN isotopic abundance}

Let us define the abundance ratio, $R_{\rm iso}$, as 
\begin{equation}
 R_{\rm iso}=\frac{^{12}{\rm C}}{^{13}{\rm C}}\frac{^{15}{\rm N}}{^{14}{\rm N}}.
\label{eq:r_iso}
\end{equation}
$R_{\rm iso}$ is found to vary with the distance from
the Galactic center (\citealt{wilson1994:araa},
\citealt{wielen1997}). The observed trend can be explained by
Galactic chemical evolution.  Other observational studies (\citealt{ikeda2002}, 
\citealt{langer1993:isotope}, \citealt{dahmen1995:isotope})
have found in addition a source-to-source variation of the isotopic ratios among a group
of clouds located at nearly the same distance from the galactic
centre.

The isotopomeric species of HCN, \htcn\ and \hcfn\ have been often
observed in their $J=$1 -- 0 transition toward various local molecular
clouds (\citealt{dahmen1995:isotope}, \citealt{hirota1998:hcn},
\citealt{ikeda2002}) to estimate the ratio $R_{\rm iso}$. These lines can be
observed with the same receiver and hence the ratio derived will be
independent of calibration errors. Moreover, the optical depth can be
measured from the three hyperfine components of the \htcn\ line. As
mentioned in Sect.~\ref{res:hcn}, \htcn\ is optically thin in almost
all cases and hence the observed ratios must indicate the abundance
ratio $R_{\rm iso}$, since the column density estimates are free from
bias due to photon trapping effects.

\begin{figure}
\centering
\includegraphics[height=\linewidth,angle=-90]{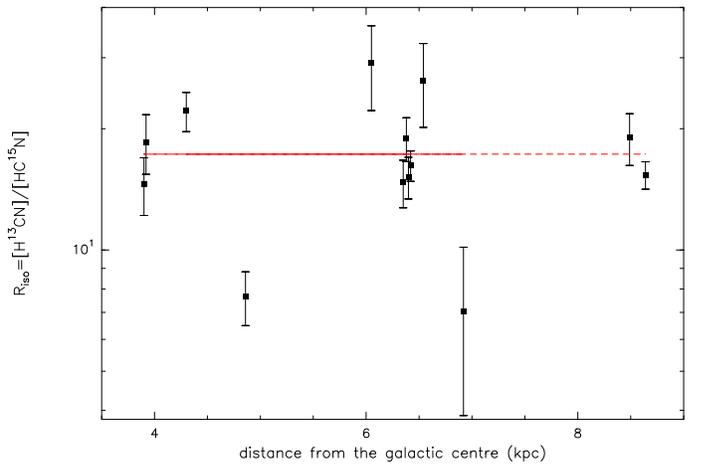}
 \caption{\htcn\ and \hcfn\ abundance ratio as a function of the distance from the galactic centre. The results are consistent with a
single value of $R_{\rm iso}=17$ shown as the dashed line.}
\label{fig:hcn}
\end{figure}

In Fig.~\ref{fig:hcn} we plot the abundance ratio of \htcn\ and \hcfn\
as a function of distance from the galactic centre for our sources.
The ratio of the abundances does not seem to correlate with the
distance.  However, the formal error bars are too large to be able to
make any strong argument against a correlation of $R_{\rm iso}$ with
galactocentric distance.

If we adopt the canonical values for the $^{12}{\rm C/^{13} C}$ of 77
and $^{14}{\rm N/^{15} N}$ of 450 toward the local interstellar medium
(ISM), then the implied \htcn\ to \hcfn\ abundance ratio is
$\sim 6$ (\citealt{wilson1994:araa}, their Table~4). From our data, we
find that most of the values tend to be higher, consistent with a mean
value of 17. This ratio is intermediate between that found toward the
galactic centre ($\sim 30$), an extreme high mass star-forming region
and the local ISM ($\sim 6$) where typically low mass star formation
pervades (see again Table~4 of \citealt{wilson1994:araa}).

  The SCAMPS sources are in the close
 vicinity of massive star forming regions and hence the energetic UV
 radiation could cause a isotope-selective photo-destruction
 \citep{schilke1992:hcn}. Chemical fractionation is also likely to
 play an important role in the observed high values. The high
 deuterium fractionation we observed must be equally affecting the
 abundances of \htcn\ and \hcfn. Therefore, if at all, chemical
 fractionation were to play a role, it must be attributed to an
 anomalous fractionation in the isotopes of either N or C. A high
 ratio implies either a low $^{15}\rm N$ fractionation or $^{13}\rm C$
 fractionation in HCN.  As noted earlier in this section, the carbon
 isotopic ratio is known to vary across the Galaxy
 \citep{wilson1994:araa}. The ratio is influenced mainly by a)
 the self-shielding property of CO, b)$^{13}\rm CO$ selective isotopic
 fractionation, and c) stellar winds from stars more massive than the sun,
 which favours the production of $^{13}\rm C$ over $^{12}\rm C$
 \citep{Kruegel2003_book}.  The deviation from the canonical value in the
 local ISM could be therefore due to a complex chemistry or
 environment in these high-mass star forming regions.

 Although the number of assumptions involved is considerable, in
 particular that the two species may not be exactly tracing the same
 volume of gas as evidenced by the differences in their line widths,
 the ratio we derive clearly indicates the differences in the
 properties of high mass and low mass star forming regions.

\section{Conclusion\label{conc}}

In this paper, we report the study of physical and chemical properties
of a new sample of high-mass pre/protocluster clumps. We have observed
32 sources in the \AMM (1,1) and (2,2), \DAMM\ (1$_{11}$--1$_{01}$),
\CO ($J=1$--0 and 2--1), \htcn\ and \hcfn\ ($J=1$--0) transitions. We have
clear detections of \DAMM\ in 22 sources and 100\% detection in
\AMM. 

Our observations suggest large deuterium enhancements (up to 66\%),
the largest reported so far.  We also derived the amount of molecular
depletion by comparing the \CO\ column density with the $\rm H_2$
column density derived from dust continuum observations, and we find that
the degree of CO depletion is a factor 2 -- 12 higher compared to the
canonical value for its abundance. The derived abundance ratio of
\htcn\ to \hcfn\ is indicative of a very high $^{13}\rm C$/$^{12}\rm
C$ ratio, discrepant from that found toward the local ISM. These
properties clearly reflect the complex chemistry in regions of high
mass star formation.

This study shows that depletion and high deuteration exist towards
massive cold cores in massive star forming regions and makes
them promising candidates for the early phases of massive star
formation. In a subsequent paper,
we will discuss the spatial distribution of \DAMM\ in two sources
observed with high angular resolution, confirming the high deuteration
found from this study.

 \appendix

\section{Radiative Transfer Equations used for Column Density Determination \label{sec:appendix}}
We used the following expressions to determine the column density of
the different molecules.
The column density is given by,
\begin{equation}
N_{\rm tot} = \frac{3 h \varepsilon_0}{2 \pi^{2} S\mu^2_g}{J(T_{\rm
ex}) Q( T_{\rm ex}) \tau \Delta v} \, ,
\label{eq:ntot_thick}
\end{equation}
where, $\Delta v$ is the linewidth, $\varepsilon_0$ is the dielectric permittivity, $S\mu^2_g$ is the line strength multiplied by the dipole moment along the molecular g-axis, 
$\tau$ is the optical depth, $h$
is the Planck constant, $T_{\rm ex}$ is the excitation temperature and
$Q(T_{\rm ex})$ is the partition function.

Here, $J(\rm T_{\rm ex})$ is defined as
 \begin{equation}
{J(\rm T_{ex}})=\frac{\exp(E_{\rm u}/kT_{\rm ex})}{\exp({h\rm \nu}/{k
T_{\rm ex}})-1} \, ,
\end{equation}
where $E_{\rm u}$ is the upper energy level, $k$ is the Boltzmann
constant and $\nu$ is the frequency of observed transition.  We assume
that the excitation temperature is the same for all rotational
levels. Equation(\ref{eq:ntot_thick}) is valid for the optically thick case.

In the optically thin case,
\begin{equation}
N_{\rm tot} = \frac{3 h \varepsilon_0}{2 \pi^{2}
S\mu^2_g}\frac{J(T_{\rm ex}) Q( T_{\rm ex}) W}{J_\nu(T_{ex})-J_\nu(
2.7)} \, ,
\label{eq:ntot_thin}
\end{equation}
where W is the integrated intensity and $J_\nu(T)$ is defined as
$J_\nu(T) = \frac{h\nu/k}{\exp({h\rm \nu}/{k T_{\rm ex}})-1}$.

The \DAMM\ partition function is determined by considering the contribution of the different energy levels from $J=0$ to $J=2$, while the metastable levels from $(J,K=1,1)$ to $(J,K=3,3)$ have been considered for \AMM.

For other molecular species, the partition function $Q(T_{\rm ex})$ at temperature $T_{\rm ex}$ is
estimated as $Q(T_{\rm ex})= \alpha T^{\beta}$, where $\alpha$ and
$\beta$ are the best fit parameters from a fit to the partition
function obtained from JPL catalogue at different excitation
temperatures from 10 -- 300~K.

In Table~\ref{tab:q_func}, we give the dipole moments we used for
different molecules and the partition function, $Q(T_{\rm ex})$ at
temperature $T_{\rm ex}=15$~K.
 
By fitting the main and the hyperfine components of the (1,1) line
 and the main component of the (2,2) line, we obtain the rotational
 temperature.  The kinetic temperature and \AMM\ column density, have
 been derived using the standard formulation for \AMM\ spectra
 (\citealt{bachiller1997:nh3}). The \AMM\ column density has been
 estimated assuming that the level populations are thermalised,
 i.e.,$T_{\rm ex}=T_{\rm rot}$. A similar assumption holds for our
 estimation of \DAMM\ column densities.

The upper limit on the \DDAMM\ column density is estimated from the
 integrated intensity. This quantity was measured by integrating over a
velocity range determined from the \DAMM\ line width around the LSR
velocity of the source. The 1$\sigma$ uncertainty on this value is
estimated from the rms of the noise in the channels where no line
emission is expected, and, is given by $\sqrt{N} v_{res} \sigma$.


\begin{table}[p{0.1cm}]
\caption{Molecular parameters used to estimate $N_{\rm tot}$ }
\small
\begin{center}
\begin{tabular}{lcc}
\hline
\hline

 Transition                         &  $S\mu^2_{\rm g}$ $^{\rm a}$       & $Q(15)$  $^{\rm b}$  \\  
                                    &   D                                &                \\    
\hline
 \DAMM\ 1$_{11}$--1$_{01}$  (para)  &    11.9                            & 47.8$^{\rm c}$   \\
 \DDAMM\ 1$_{10}$--1$_{01}$ (ortho) &    0.722                           & 50.2 $^{\rm d}$    \\
 \htcn\  $J=1$--0, $F=2$--1         &    14.84                           & 22.8$^{\rm e}$     \\
 \hcfn\  $J=1$--0                   &    8.909                           & 7.6 $^{\rm e}$     \\
 \CO\    $J=1$--0                   &    0.012                           &  6.1$^{\rm e}$     \\
 \CO\    $J=2$--1                   &    0.024                           & 6.1$^{\rm e}$      \\ 
 \hline
\end{tabular}
\end{center}
$^{\rm a}$ S is the line strength and $\mu_{\rm g}$  is the dipole moment along the molecular g-axis.\\
$^{\rm b}$ the partition function corresponding to a temperature of 15~K, the typical temperature of the sources in the present sample. \\
$^{\rm c}$ the partition function obtained by the summation over the contribution of the different energy levels from $J=0$ to $J=2$. \\
$^{\rm d}$ the partition function based on data from the CDMS catalog \citep{mueller2005:cdms}. \\
$^{\rm e}$ the partition function based on data from the JPL catalog \citep{pickett1998:jpl}. \\
\label{tab:q_func}
\end{table}

\acknowledgements{ T.Pillai was
supported for this research through a stipend from the International
Max Planck Research School (IMPRS) for Radio and Infrared Astronomy at
the University of Bonn. JH was supported at MPIfR Bonn by DFG SFB 494
and holds a PPARC AF at Exeter. 
}

\bibliographystyle{aa}
\bibliography{bib_astro}
 
\Online

\onltab{2}{
\begin{table*}[h]
\begin{center}
\caption{\AMM\ line parameters with uncertainties (in brackets) from the hyperfine/Gaussian fits.}
\vspace{1em}
\begin{tabular}{|l|ccc|c|}
\hline
& \multicolumn{3}{c|}{\AMM\ (1,1)} & \multicolumn{1}{c|}{\AMM\ (2,2)} \\
Source  & \multicolumn{1}{c}{$T_{\rm MB}$$^{\rm a}$ } & \multicolumn{1}{c}{$\Delta$v$^{\rm b}$} &
\multicolumn{1}{c|}{$\tau_{\rm mg}$$^{\rm c}$} & \multicolumn{1}{c|}{$T_{\rm MB}$} \\\hline
        & \multicolumn{1}{c}{(K)} & \multicolumn{1}{c}{\kms} & \multicolumn{1}{c|}{} & \multicolumn{1}{c|}{(K)} \\\hline
G8.13+0.25      & 1.9   (0.4)  &         1.5   (0.1) &  1.5    (0.3)   & 1.0   (0.2) \\
                & 2.4   (0.4)  &         1.3   (0.1) &  1.9    (0.3)   & 1.3   (0.2) \\
G8.68-0.37      & 5.3   (0.5)  &         1.1   (0.0) &  5.0    (0.4)   & 2.3   (0.3) \\
                & 5.9   (0.4)  &         3.0   (0.1) &  2.6    (0.1)   & 3.5   (0.2) \\
G8.71-0.37      & 3.5   (0.4)  &         1.3   (0.0) &  4.2    (0.3)   & 1.7   (0.2) \\
G10.15-0.34     & 1.1   (0.1)  &         4.1   (0.3) &  1.1    (0.5)   & 0.7   (0.1) \\
G10.21-0.31     & 2.9   (0.4)  &         1.9   (0.0) &  3.9    (0.3)   & 1.8   (0.1) \\
G10.21-0.32     & 3.4   (0.5)  &         1.7   (0.0) &  3.5    (0.2)   & 1.8   (0.3) \\
                & 4.6   (0.4)  &         1.9   (0.0) &  4.6    (0.2)   & 2.7   (0.1) \\
G10.61-0.33     & 2.0   (0.2)  &         1.9   (0.1) &  1.8    (0.3)   & 1.1   (0.1) \\
G11.11-0.12P1   & 5.1   (1.0)  &         1.3   (0.1) &  3.5    (0.4)   & 1.9   (0.3) \\
G11.11-0.12P3   & 3.0   (2.2)  &         2.5   (0.1) &  2.4    (0.2)   & 1.0   (0.2) \\
G11.11-0.12P4   & 4.3   (0.9)  &         1.5   (0.1) &  3.1    (0.6)   & 0.9   (0.4) \\
G12.19-0.12     & 0.4   (0.0)  &         3.9   (0.2) &  1.8    (0.3)   & 0.1   (0.0) \\
G13.18+0.06     & 5.4   (0.5)  &         2.4   (0.0) &  3.3    (0.1)   & 3.7   (0.1) \\
G15.01-0.67     & 5.0   (0.4)  &         2.9   (0.1) &  0.7    (0.1)   & 3.6   (0.1) \\
G15.01-0.69     & 3.3   (0.2)  &         2.6   (0.1) &  0.5    (0.1)   & 2.3   (0.1) \\
G15.03-0.65     & 2.6   (0.3)  &         2.4   (0.2) &  1.2    (0.3)   & 1.6   (0.2) \\
                & 2.8   (0.3)  &         2.6   (0.1) &  1.3    (0.3)   & 1.2   (0.4) \\
G18.17-0.30     & 2.9   (0.3)  &         1.6   (0.0) &  3.3    (0.2)   & 1.7   (0.1) \\
G18.21-0.34     & 3.8   (0.4)  &         1.5   (0.0) &  2.8    (0.2)   & 2.0   (0.2) \\
G19.30+0.07P1   & 5.0   (0.7)  &         2.0   (0.1) &  2.1    (0.3)   & 2.7   (0.2) \\
G19.30+0.07P2   & 3.2   (0.6)  &         1.7   (0.1) &  2.9    (0.5)   & 1.4   (0.3) \\
G23.41-0.23     & 3.0   (0.3)  &         2.0   (0.1) &  1.2    (0.1)   & 1.6   (0.2) \\
G23.42-0.23     & 2.2   (0.3)  &         1.9   (0.1) &  1.6    (0.2)   & 1.3   (0.1) \\
                & 1.3   (0.4)  &         1.5   (0.1) &  0.6    (0.4)   & 0.4   (0.1) \\
G23.44-0.18     & 3.6   (0.2)  &         3.6   (0.1) &  1.7    (0.1)   & 2.8   (0.1) \\
G27.29+0.15     & 1.7   (0.2)  &         2.5   (0.1) &  1.6    (0.2)   & 0.9   (0.1) \\
G27.31+0.18     & 1.0   (0.2)  &         2.1   (0.1) &  2.1    (0.4)   & 0.6   (0.1) \\
G28.34+0.06P1   & 2.7   (0.5)  &         2.7   (0.0) &  2.0    (0.1)   & 1.3   (0.1) \\
G28.34+0.06P2   & 2.8   (0.5)  &         2.0   (0.1) &  2.7    (0.3)   & 1.5   (0.2) \\
G29.97-0.05     & 0.8   (0.2)  &         2.0   (0.0) &  2.4    (0.2)   & 0.3   (0.0) \\
G33.71-0.01     & 2.0   (0.4)  &         2.4   (0.2) &  2.9    (0.6)   & 1.5   (0.2) \\
G34.81-0.28     & 4.7   (0.5)  &         1.4   (0.0) &  1.6    (0.2)   & 2.2   (0.1) \\
G35.19-1.73     & 4.7   (0.4)  &         1.6   (0.0) &  1.8    (0.1)   & 2.4   (0.1) \\
G79.34+0.33     & 2.4   (0.3)  &         1.2   (0.0) &  1.5    (0.1)   & 0.7   (0.1) \\
G81.74+0.59     & 6.1   (0.6)  &         2.2   (0.0) &  1.3    (0.1)   & 3.2   (0.1) \\

\hline
\end{tabular}

\label{tab:nh3_line parameter}
\end{center}
$^{\rm a}$ Main beam brightness temperature from gaussian fit. \\
$^{\rm b}$ FWHM from the hyperfine fits. \\
$^{\rm c}$ optical depth of the main group of hyperfines from the hyperfine fits. \\
Note: Five sources clearly have at least 2 velocity components, hence the fit parameters have been determined for each component. Similar notations are used for other molecules elsewhere in this paper. The parameter has been fixed for the fit for those entries with no error indicated.
\end{table*}
}

\onltab{3}{
\begin{table*}[h]
\begin{center}
\caption{\DAMM\ line parameters with uncertainties (in brackets) from the hyperfine/Gaussian fits.}
\vspace{1em}
\begin{tabular}{|l|ccc|}
\hline
& \multicolumn{3}{c|}{\DAMM\ 85.93~GHz} \\
Source  & \multicolumn{1}{c}{\tdv$^{\rm a}$} & \multicolumn{1}{c}{$\Delta$v$^{\rm b}$} & \multicolumn{1}{c|}{$\tau_{\rm tot}$$^{\rm c}$ } \\\hline
        & \multicolumn{1}{c}{K~\kms} & \multicolumn{1}{c}{\kms} & \multicolumn{1}{c|}{ } \\\hline
G8.13+0.25      & 0.5     ($<$0.1)  & 1.3     (0.3 )    & --             \\
                & 0.7     ($<$0.1)  & 1.8     (0.9 )    & --             \\
G8.68-0.37      & --                & 1.2     (0.1 )    & 2.2    ( 0.7 ) \\
G8.71-0.37      & --                & 1.1     ($<$0.1)  & 3.7    ( 0.6 ) \\
G10.21-0.31     & --                & 1.2     (0.1 )    & 3.4    ( 1.4 ) \\
G10.21-0.32     & --                & 1.4     (0.1 )    & 2.3    ( 1.0 ) \\
G10.61-0.33     & 0.3     ($<$0.1)  & 1.5     (0.2 )    & --             \\
G11.11-0.12P1   & --                & 1.0     (0.1 )    & 3.6    ( 1.5 ) \\
G11.11-0.12P4   & 0.4     (0.1 )    & 1.3     (0.2 )    & --             \\
G12.19-0.12     & 1.5     (0.1 )    & 3.6     (0.5 )    & --             \\
G13.18+0.06     & --                & 2.2     (0.1 )    & 2.0    ( 0.5 ) \\
G18.17-0.30     & --                & 1.5     (0.1 )    & 4.4    ( 1.1 ) \\
G18.21-0.34     & --                & 1.3     (0.1 )    & 2.0    ( 0.8 ) \\
G23.41-0.23     & --                & 1.7     (0.2 )    & 2.1    ( 1.3 ) \\
G23.44-0.18     & 1.4     (0.1 )    & 3.1     (0.2 )    & --             \\
G27.29+0.15     & 0.4     (0.1 )    & 1.5     (0.4 )    & --             \\
G28.34+0.06P2   & 1.2     (0.1 )    & 1.8     (0.2 )    & --             \\
G29.97-0.05     & --                & 1.4     (0.1 )    & 3.0    ( 0.6 ) \\
G33.71-0.01     & 0.4     ($<$0.1)  & 1.6     (0.3 )    & --             \\
G34.81-0.28     & 0.8     (0.1 )    & 1.7     (0.1 )    & --             \\
G35.19-1.73     & --                & 1.4     ($<$0.1)  & 2.7    ( 0.4 ) \\
G79.34+0.33     & --                & 1.0     (0.9 )    & 1.1    ( 0.1 ) \\
G81.74+0.59     & --                & 2.8     (0.1 )    & 1.5    ( 0.3 ) \\

\hline
\end{tabular}

\label{tab:nh2d_line parameter}
\end{center}

$^{\rm a}$ Integrated intensity of the line from the Gaussian fits. \\
$^{\rm b}$ FWHM  from the hyperfine fits. \\
$^{\rm c}$ Total optical depth from the hyperfine fits. \\
Note: Similar notations are used for other molecules elsewhere in this paper. The parameter has been fixed for the fit for those entries with no error indicated.
\end{table*}
}


\onlfig{1}{
\begin{figure*}
\centering
 \includegraphics[height=\linewidth,angle=0]{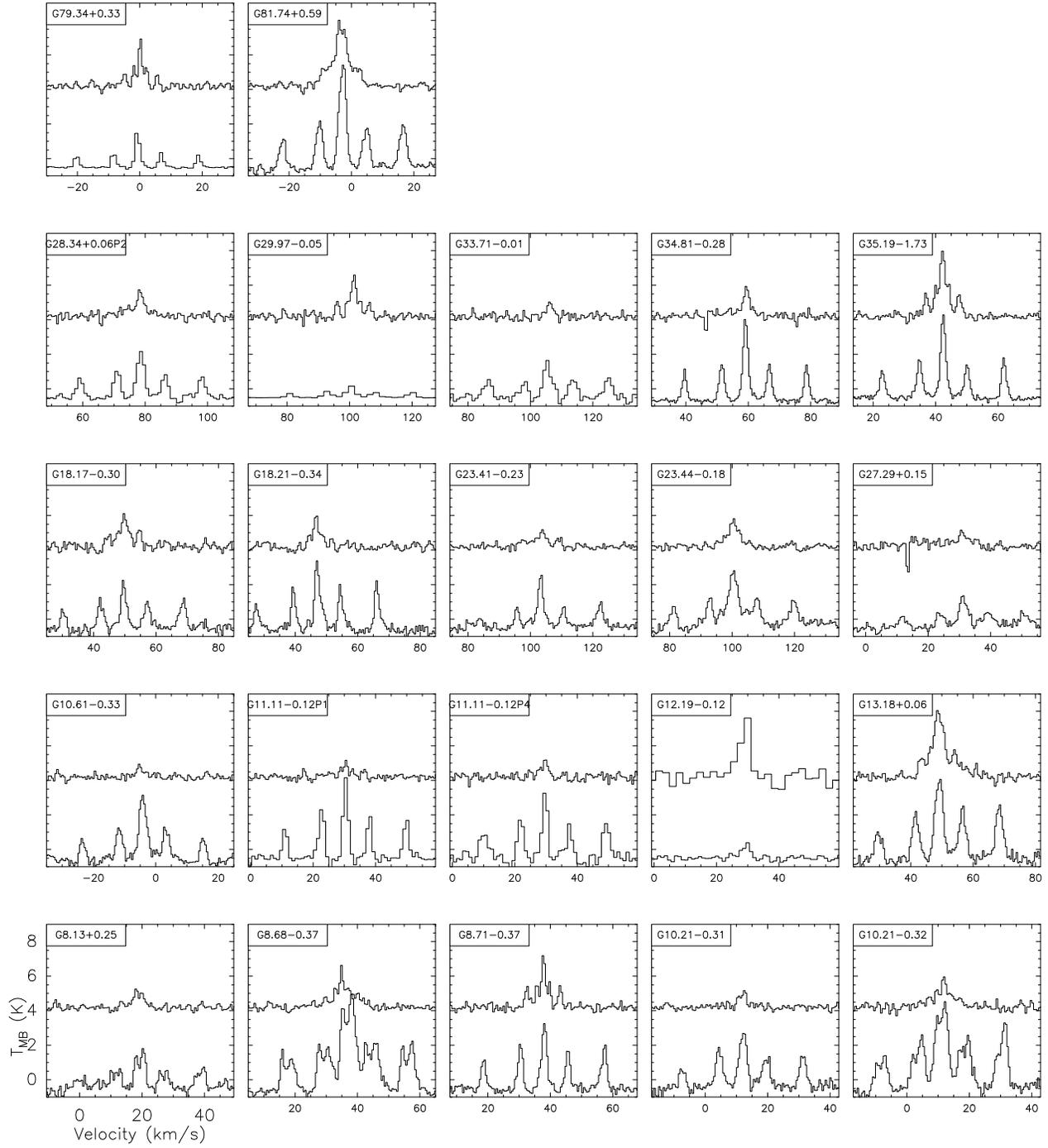}
 \caption{\textit{Lower spectrum in each panel}: Effelsberg 100m observation of the \AMM\
   (1,1) emission.\textit{Upper spectrum in each panel}: 30m observation of the \DAMM\ at
   85.9~GHz. The \DAMM\ is scaled by a factor 5.0 to amplify the
   emission relative to \AMM\ in the absolute (brightness temperature) units.}
\label{fig:nh2d_nh3}
\end{figure*}
}

\onlfig{2}{
\begin{figure*}
\centering
 \includegraphics[height=\linewidth,angle=90]{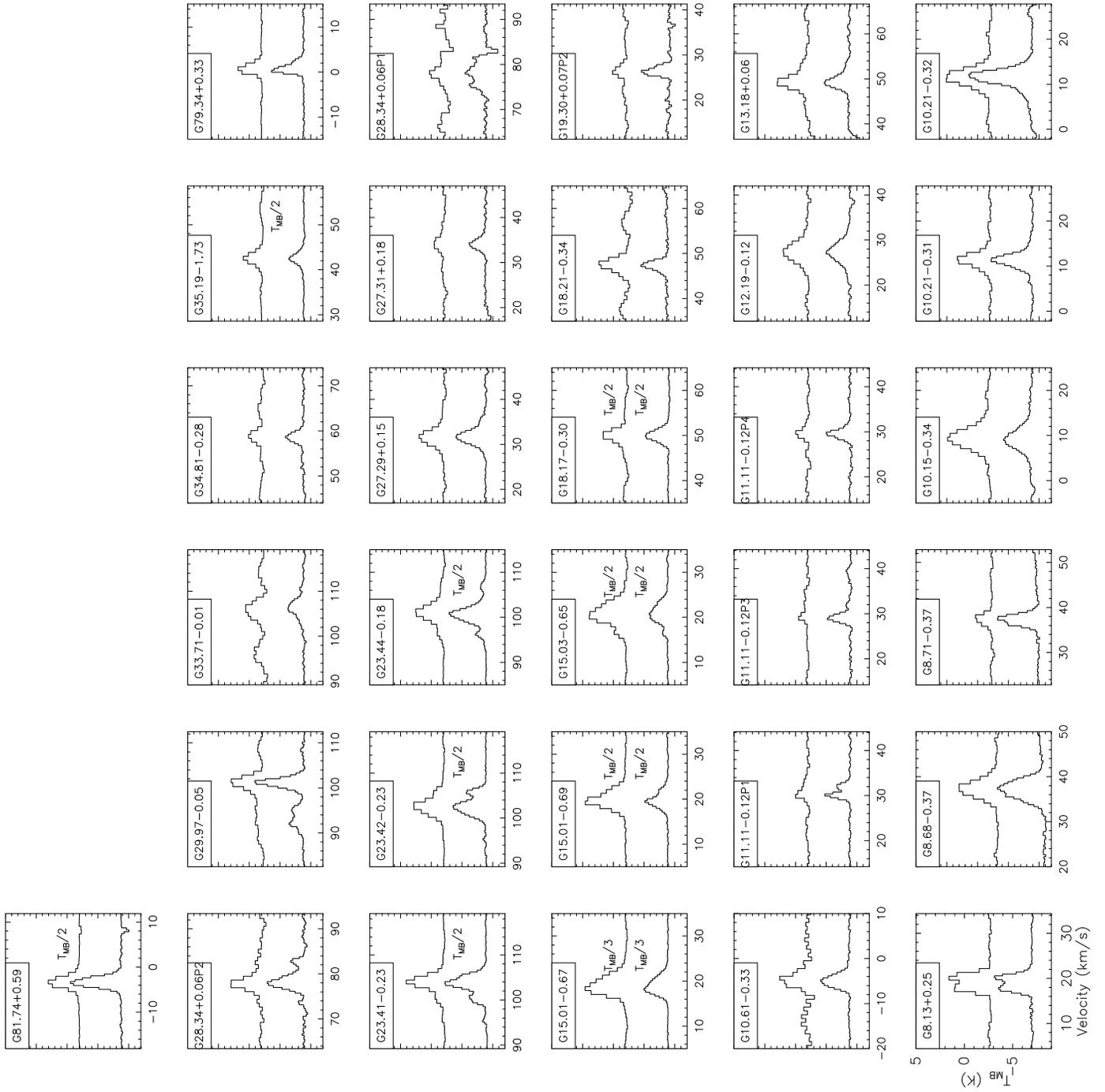}
 \caption{lower spectrum:30m \CO\ ($J=1$--0). upper spectrum: 30m \CO\ ($J=2$--1). The velocity range is $\pm$15~\kms\ of the systemic velocity.}
\label{fig:c18o_spec}
\end{figure*}
}

\onltab{4}{
\begin{table*}[p{0.1cm}]
\begin{center}
\caption{\CO\ ($J=1$--0) and ($J=2$--1) line parameters }
\vspace{1em}
\begin{tabular}{|l|c|c|}
\hline
        & \multicolumn{1}{c|}{ \CO\  ($J=1$--0)} & \multicolumn{1}{c|}{\CO\  ($J=2$--1) } \\
Source  & \multicolumn{1}{c|}{\tdv}    & \multicolumn{1}{c|}{\tdv}   \\
\hline
        & \multicolumn{1}{c|}{K~\kms}  & \multicolumn{1}{c|}{K~\kms}  \\
\hline
G8.13+0.25                & 15.7  (0.1) & 27.6  (0.5)           \\
G8.68-0.37                & 24.9  (0.9) & 31.9  (1.7)           \\
G8.71-0.37                & 13.5  (0.4) &  5.9  (0.7)           \\
G10.15-0.34               & 15.6  (0.3) & 37.4  (0.5)           \\
G10.21-0.31               & 14.9  (0.2) & 19.0  (0.4)           \\
G10.21-0.32               & 27.5  (0.3) & 33.2  (0.6)           \\
G10.61-0.33               & 10.6  (0.3) & 14.7  (0.5)           \\
G11.11-0.12P1             & 6.1   (0.1) &  5.5  (0.4)           \\
G11.11-0.12P3             & 5.1   (0.3) &  5.1  (0.5)           \\
G11.11-0.12P4             & 5.1   (0.2) &  4.8  (0.5)           \\
G12.19-0.12               & 12.0  (0.2) & 20.8  (0.4)           \\
G13.18+0.06               & 9.7   (0.5) & 20.9  (0.8)           \\
G15.01-0.67               & 31.0  (0.2) & 91.7  (1.0)           \\
G15.01-0.69               & 16.2  (0.2) & 53.7  (0.5)           \\
G15.03-0.65               & 19.8  (0.2) & 65.5  (0.6)           \\
G18.17-0.30               & 13.1  (0.7) & 26.5  (1.3)           \\
G18.21-0.34               & 7.3   (0.4) & 12.0  (1.3)           \\     
G19.30+0.07P1$^{\rm a}$   & 1.1   (0.4) &  --         \\
G19.30+0.07P2   & 4.1   (0.4) &  6.6  (0.7)           \\
G23.41-0.23     & 35.2  (0.8) & 44.5  (1.0)           \\
G23.42-0.23     & 31.5  (0.3) & 43.1  (0.5)           \\
G23.44-0.18     & 31.1  (0.5) & 43.4  (0.6)           \\
G27.29+0.15     & 12.4  (0.2) & 16.6  (0.5)           \\
G27.31+0.18     & 4.9   (0.2) &  6.1  (0.5)           \\
G28.34+0.06P1   & 11.3  (0.4) &  6.3  (1.5)           \\
G28.34+0.06P2   & 14.4  (0.5) & 19.1  (0.7)           \\
G29.97-0.05     & 12.9  (0.7) & 18.2  (0.9)           \\
G33.71-0.01     & 7.4   (0.2) &  9.0  (1.3)           \\
G34.81-0.28     & 5.4   (0.3) &  6.6  (0.6)           \\
G35.19-1.73     & 8.4   (0.2) & 15.8  (0.8)           \\
G79.34+0.33     & 7.8   (0.1) &  9.9  (0.2)           \\
G81.74+0.59     & 14.9  (0.2) & 32.4  (0.4)           \\        
\hline
\end{tabular}
\label{tab:c18o_line_parameter}
\end{center}
\tdv\ is the integrated intensity over $\pm$5~\kms of the brightest \CO\ ($J=1$--0) peak.\\
$^{\rm a}$ \CO\ ($J=2$--1) data suffered from baseline ripples.

\end{table*}
}

\onlfig{3}{
\begin{figure*}

\centering
 \includegraphics[height=0.9\linewidth,angle=0]{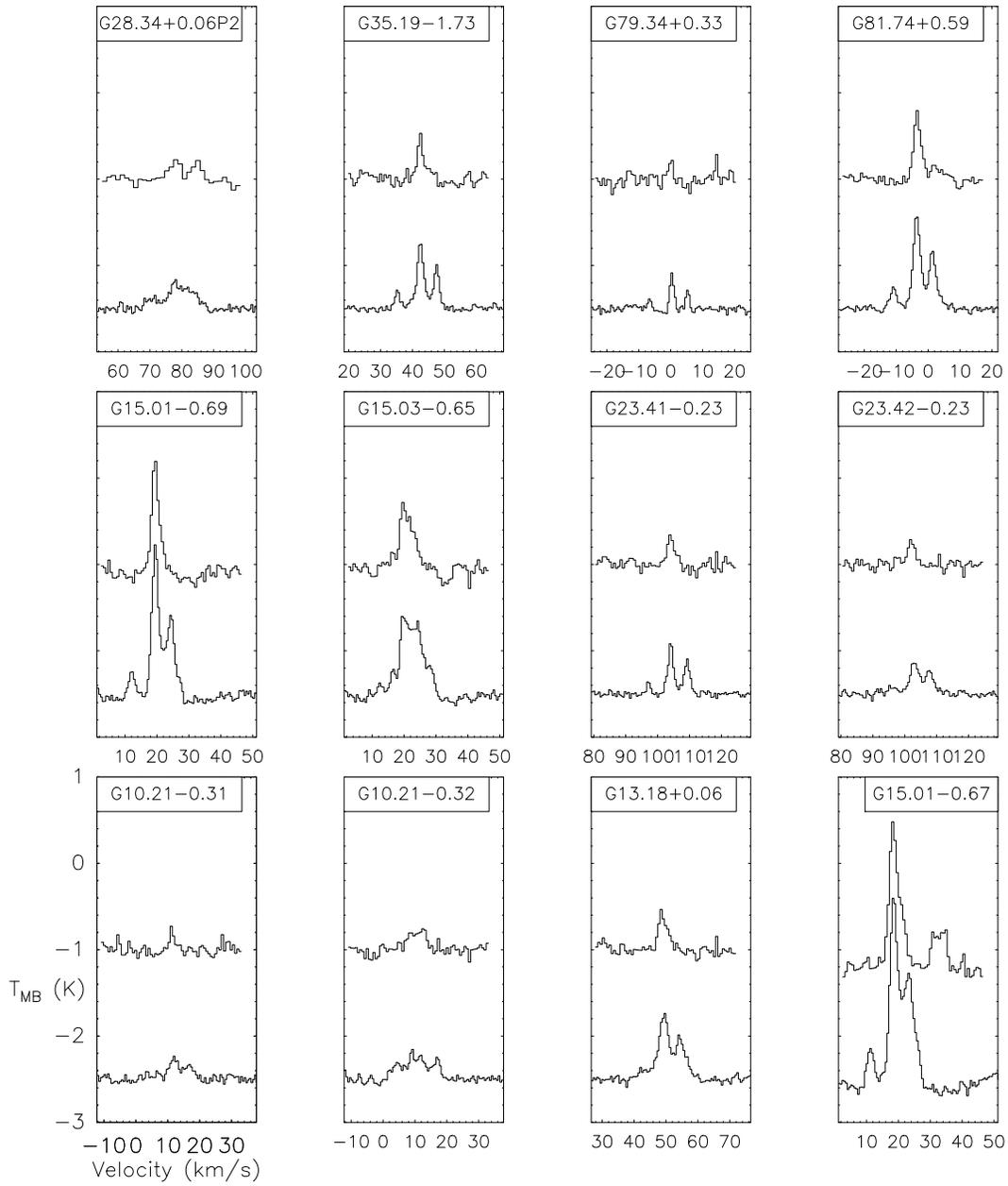}
 \caption{lower spectrum: 30m observation of \htcn\ . upper spectrum: 30m observation of \hcfn. The \hcfn\ is scaled by a factor 2 to amplify the emission relative to \htcn\ in the absolute units. The velocity range is $\pm$25~\kms\ of the systemic velocity.}
\label{fig:h13cn_hc15n}
\end{figure*}
}

\end{document}